\documentclass[12pt,preprint]{aastex}
\usepackage{amssymb, amsmath}
\usepackage{color}
\usepackage{morefloats}
\usepackage{hyperref}

\shortauthors{Neugent et al.}

\begin{document}

\title{Binary Red Supergiants II: \\Discovering and Characterizing B-type Companions}

\author{Kathryn F.\ Neugent\altaffilmark{1,2}, Emily M.\ Levesque\altaffilmark{1}, Philip Massey\altaffilmark{2,3}, and Nidia I.\ Morrell\altaffilmark{4}}

\altaffiltext{1}{Department of Astronomy, University of Washington, Seattle, WA, 98195 ; kneugent@uw.edu; emsque@uw.edu}
\altaffiltext{2}{Lowell Observatory, 1400 W Mars Hill Road, Flagstaff, AZ 86001; phil.massey@lowell.edu}
\altaffiltext{3}{Department of Physics and Astronomy, Northern Arizona University, Flagstaff, AZ, 86011-6010}
\altaffiltext{4}{Las Campanas Observatory, Carnegie Observatories, Casilla 601, La Serena, Chile; nmorrell@lco.cl}

\begin{abstract}
The percentage of massive main sequence OB stars in binary systems is thought to be as high as 100\%. However, very few Galactic binary red supergiants (RSGs) have been identified, despite the fact that these stars are the evolved descendants of OB stars. As shown in our recent paper, binary RSGs will likely have B-type companions, as dictated by stellar evolution considerations. Such a system will have a very unique photometric signature due to the shape of the spectral energy distribution. Using photometric cutoffs it should therefore be possible to detect candidate RSG+B star binary systems. Here we present our spectoscopic follow-up observations of such candidates. Out of our initial list of 280 candidates in M31 and M33, we observed 149 and confirmed 63 as newly discovered RSG+B star binary systems. Additional spectra of four candidate systems in the Small Magellanic Cloud confirmed all of them as new RSG+B star binaries including the first known RSG+Be star system. By fitting BSTAR06 and MARCS model atmospheres to the newly-obtained spectra we place estimates on the temperatures and subtypes of both the B stars and RSGs. Overall, we have found 87 new RSG+B star binary systems in M31, M33 and the Small and Large Magellanic Clouds. Our future studies are aimed at determining the binary fraction of RSGs.
\end{abstract}

\section{Introduction}
Characterizing the stellar binary population is one of the keys to understanding stellar evolution. At close enough binary separations, companions disrupt circumstellar environments and strip off outer layers through induced mass transfer. This stripping removes the lighter elements from the surface and reveals the heavier elements, causing changes in observable properties, and greatly changes the evolutionary path (see upcoming proceedings of the ESO conference {\it The Impact of Binary Stars on Stellar Evolution}.). Despite its importance, the binary fraction of many types of stars, especially massive stars, is still poorly understood. Here we continue our investigation of the binary population of massive red supergiants (RSGs).

The observed short-period binary fraction of un-evolved massive stars (O-type stars) is 30-40\% \citep{Garmany80,Sana30Dor}, with a similar fraction found for their high-mass evolved descendants, the Wolf-Rayet (WR) stars \citep{vanDerHucht2000, Foellmi2003, WRbins}. However, when longer period systems are included, and selection effects are accounted for, estimates of the fraction of O-type stars with companions approaches 100\% \citep{Gies08, SanaSci}. Until recently, not much was known about the binarity of red supergiants (RSGs) which descend from 9-30$\rm{M_\odot}$ OB stars. Given the high binary fraction of un-evolved O-type stars, it follows that there should also be a high binary fraction of RSGs, but until a year ago the total number of known Galactic binary RSGs was 11, as listed in Table~\ref{tab:knownBinaries}. {\it Where were all the missing binaries?}

\citet{NeugentRSG} launched both a theoretical and observational investigation of this important question. They developed a set of photometric cutoffs that identify binary RSGs. This was made easier by discovering that, from an evolutionary point of view, RSGs preferentially exist in binary systems with B-type stars. Later type dwarfs (A,F,G,K, and M) will not yet have had time to form. Higher mass companions (RSG+WRs, for example) will be rare given the short lifetimes of such stars. Since the spectral energy distributions (SEDs) of B-type stars peak in the UV and the SEDs of RSGs peak in the IR, a combination of these two SEDs will look different from either a B-type star or a RSG by themself. \citet{NeugentRSG} created a grid of $\sim$25,000 synthetic spectral models by combining synthetic B-type spectra from the BSTAR06 models \citep{BSTARS06} with RSG synthetic spectra from the MARCS models \citep{MARCS}. U-B and R-I photometry of these models led to a set of color-color cutoffs that could be used to photometrically select candidate RSG+B star binaries.

Using the Local Group Galaxy Survey (LGGS) \citep{LGGS} for stars in M31 and M33, \citet{NeugentRSG} came up with a list of 280 candidate RSG+B star binaries based on their U-B and R-I colors. A subset of these candidates have now been observed for spectroscopic confirmation. This paper details the results of these observations, as well as the analysis of their characteristics.  In addition, we obtained follow-up observations of four SMC RSGs discussed in \citet{NeugentRSG} which had insufficient signal-to-noise (S/N) to determine whether or not they were binaries; we confirm all four here. Finally we conclude by laying out the path to determining the RSG binary frequency.

\section{Discovery of New RSG+B Star Binaries in M31 and M33}
\citet{NeugentRSG} selected 592 candidate RSG+B star binaries from the LGGS for M31 and M33. The candidates were initially selected through color-color cutoffs before then being filtered through {\it Gaia} DR2 \citep{Gaia} to identify and remove 312 foreground sources. The remaining 280 stars (138 in M31 and 142 in M33) became our initial list of candidates. 

\subsection{Observations and Reductions}
To observe as many of the 280 stars as possible, we relied on the multi-object fiber-fed spectrograph, Hectospec, \citep{Fabricant2005} on the 6.5-m MMT. Its large field of view (1 degree in diameter) was well matched to our M31 and M33 survey areas. We were assigned 2.5 nights of dark time in Fall 2018 through the Arizona Time Allocation Committee in order to study a number of WR+O binaries in M31 and M33, but since Hectospec has 300 fibers, we were able to piggyback the RSG binarity project onto this time using unused fibers. Due to poor weather and instrument problems, only a single M31 and a single M33 configuration were observed. The M31 field was observed on UT 2018 September 14 while the M33 field was observed on UT 2018 October 6 both with clear sky conditions. The data were taken with the 270 line mm$^{-1}$ grating resulting in spectral coverage from 3700 - 9000 \AA, ideal coverage for both the upper Balmer lines (3700\AA\ - 4000\AA) originating from the B-type stars and the TiO bands (upwards of 6000\AA) of the RSGs. The 250 $\mu$m fibers resulted in a spectral resolution of 6\AA. Exposures were $3\times3000$ seconds. Reductions were carried out using the version 2.0 of the Hectospec REDuction (HSRED) code\footnote{See, \url{https://www.cfa.harvard.edu/mmti/hectospec/hecto_pipe_report.pdf} and \url{https://www.mmto.org/node/536}.} which is similar to the SDSS spectral pipeline. The data were then flux calibrated using an approximate sensitivity function kindly provided by Nelson Caldwell.

\subsection{New Binaries}
\begin{figure}
\epsscale{1}
\plotone{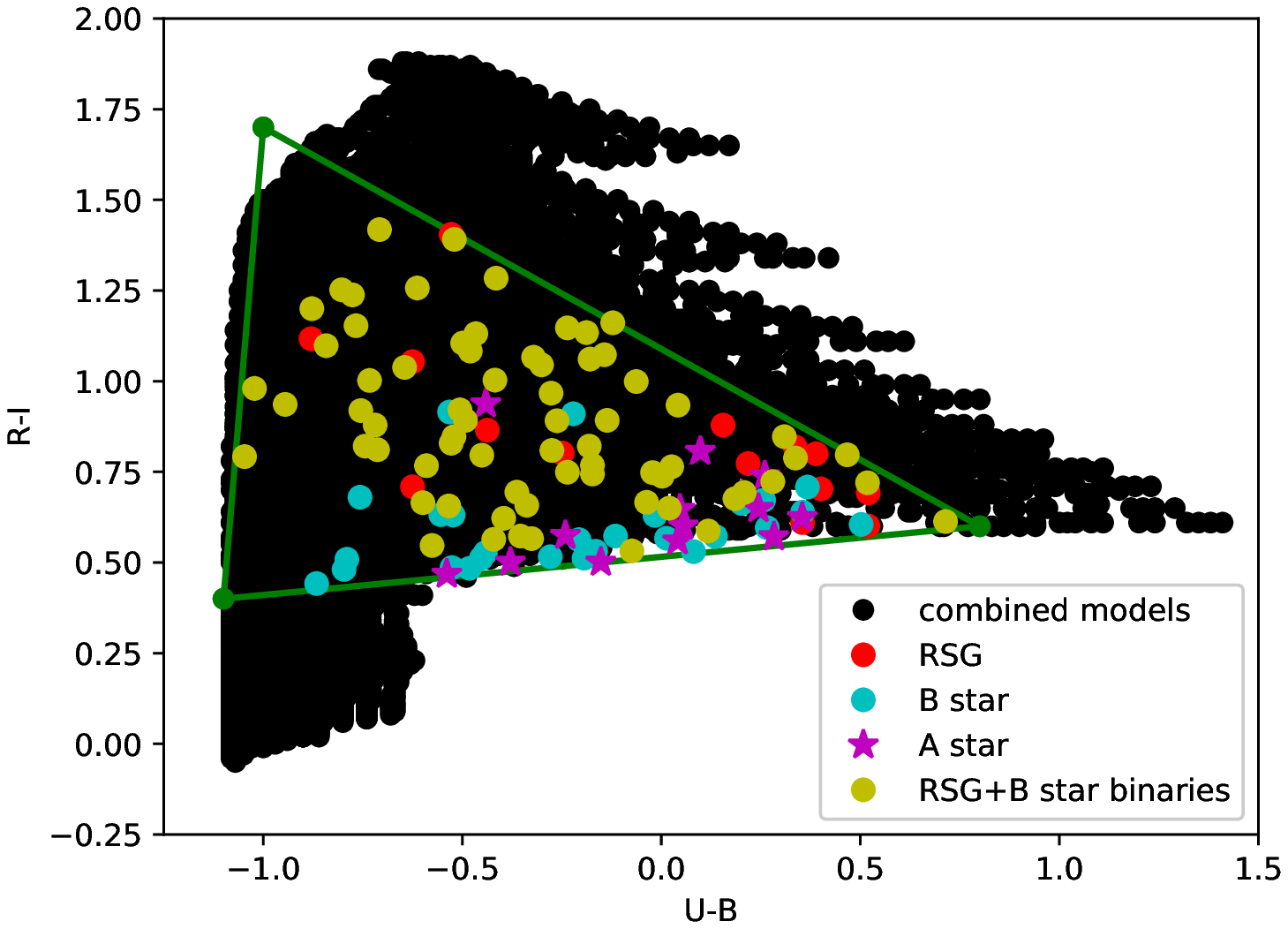}
\caption{\label{fig:newDiscovered} Location of observed RSG+B star binary candidates in R-I vs.\ U-B color-color space. The confirmed RSG+B star binaries showing both TiO bands and upper Balmer lines are shown as yellow dots. The RSGs with just TiO bands are shown as red dots. The B stars with just upper Balmer lines are shown as blue dots. The A stars are shown as magenta stars. The outline of our photometric cutoffs is shown as a green triangle. Note that the RSG+B star binaries are spread without the entire region in color-color space while the bluer B and A type stars are, for the most part, confined to the low R-I values. Some of the single RSGs are closer to the redder portion in color-color space but half of the sample shows evidence of a blue component in the spectrum.}
\end{figure}

\begin{figure}
\epsscale{0.45}
\plotone{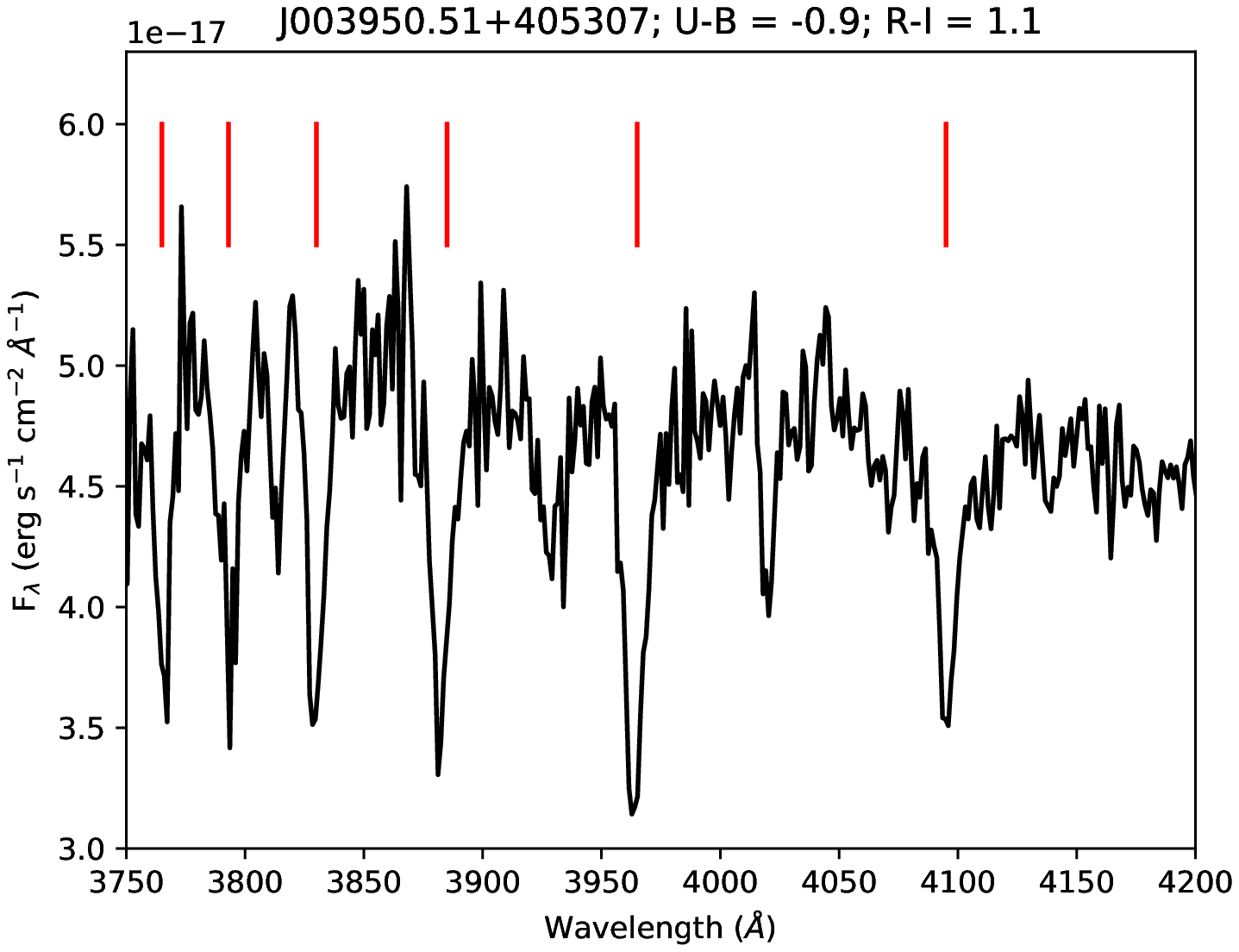}
\plotone{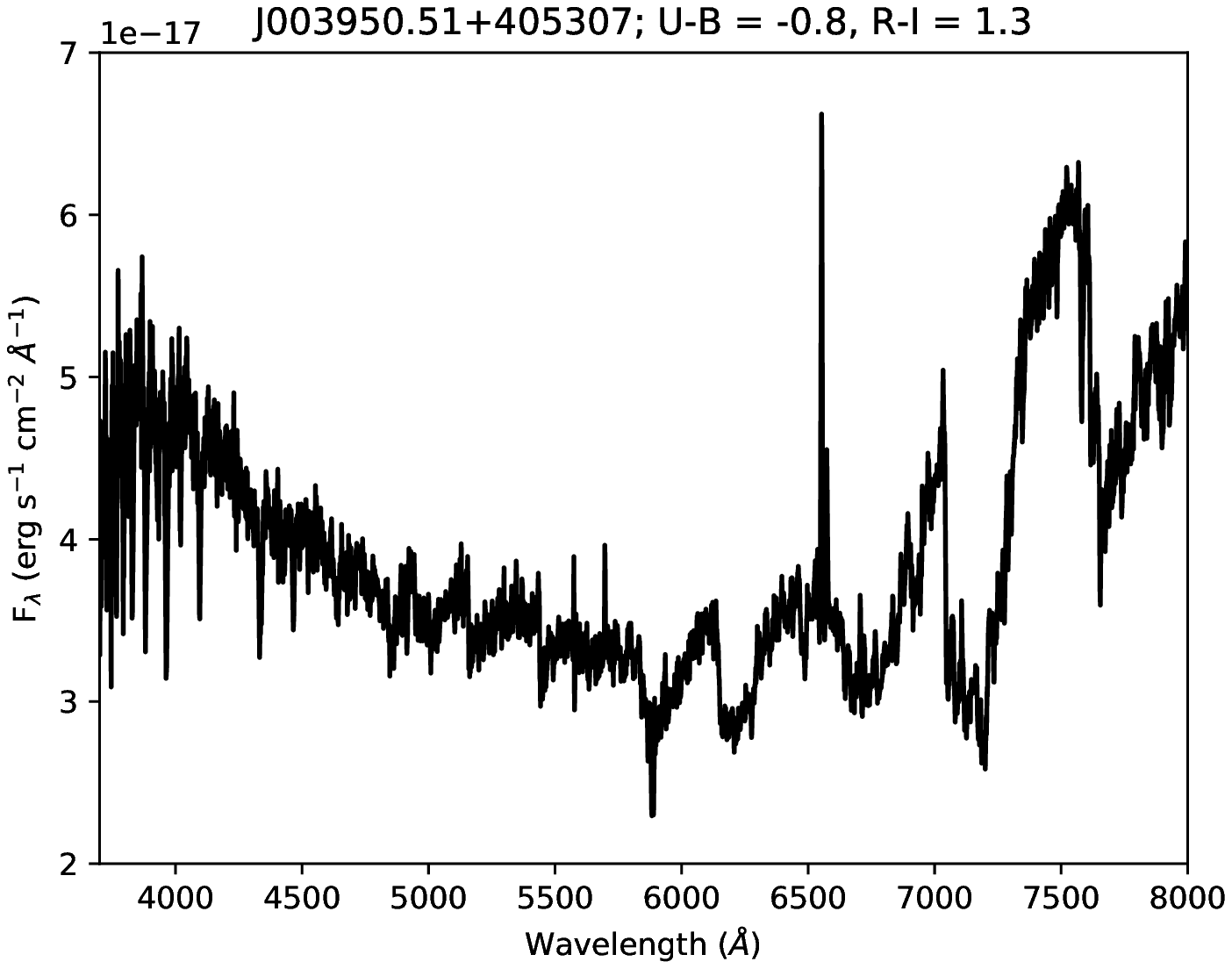}
\plotone{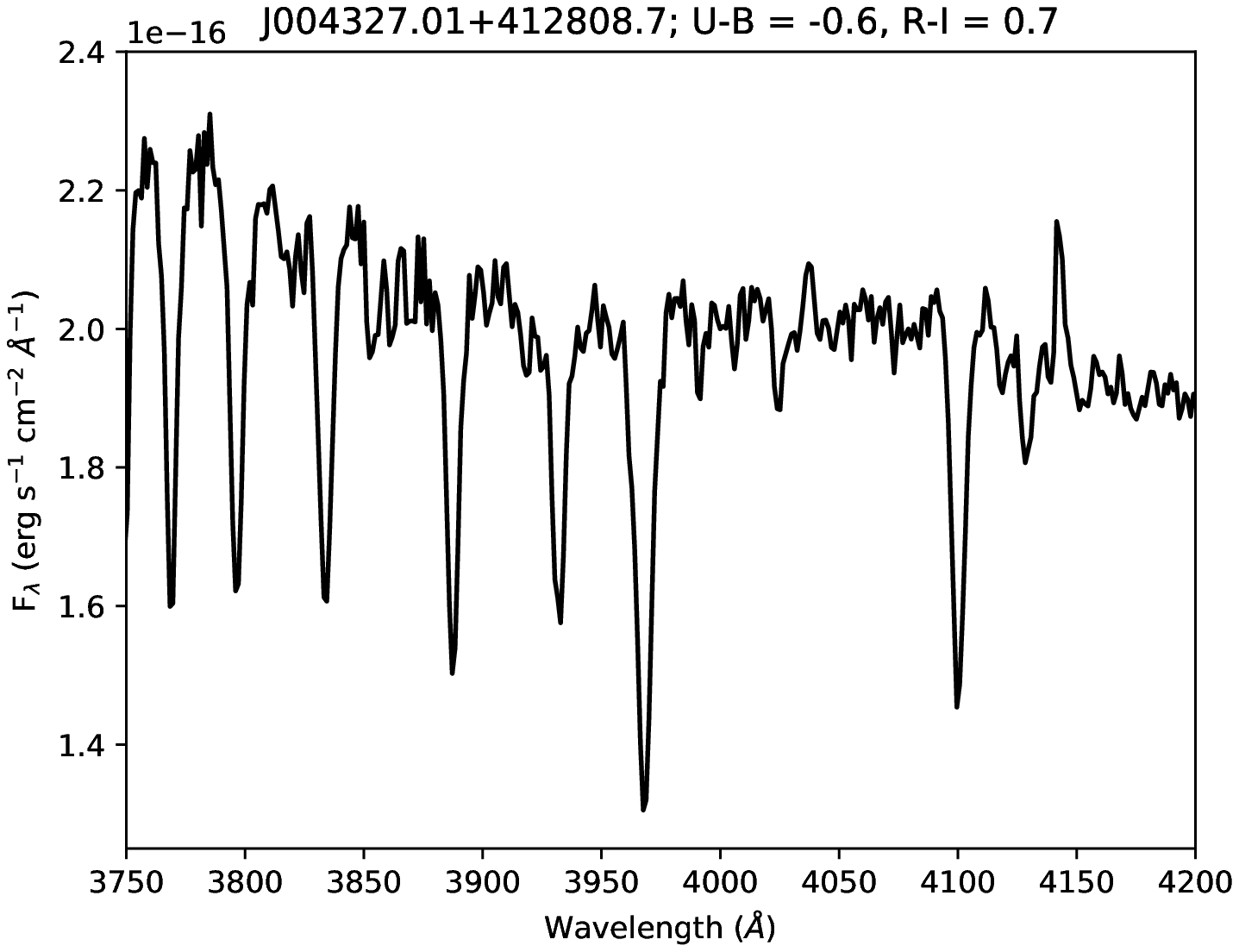}
\plotone{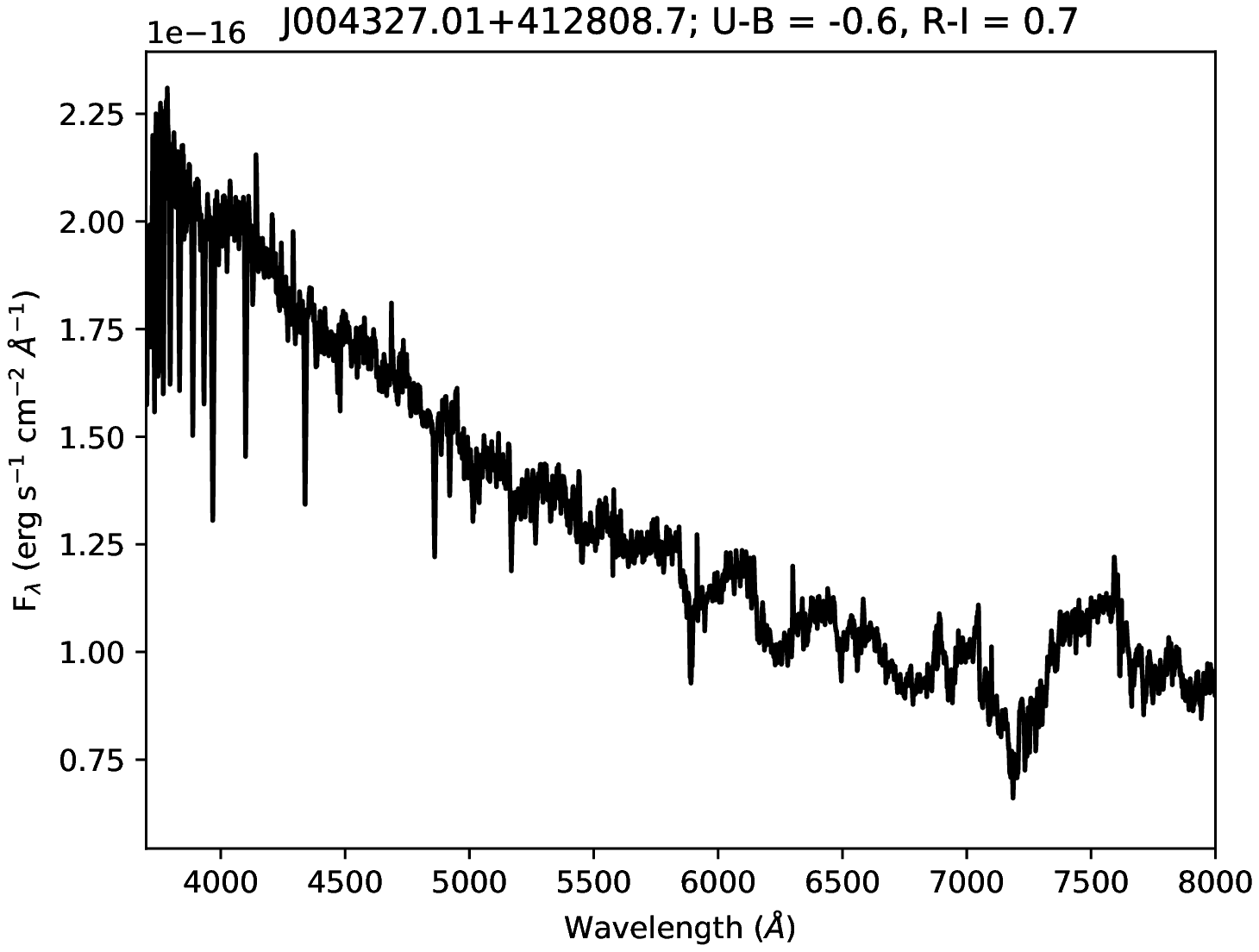}
\plotone{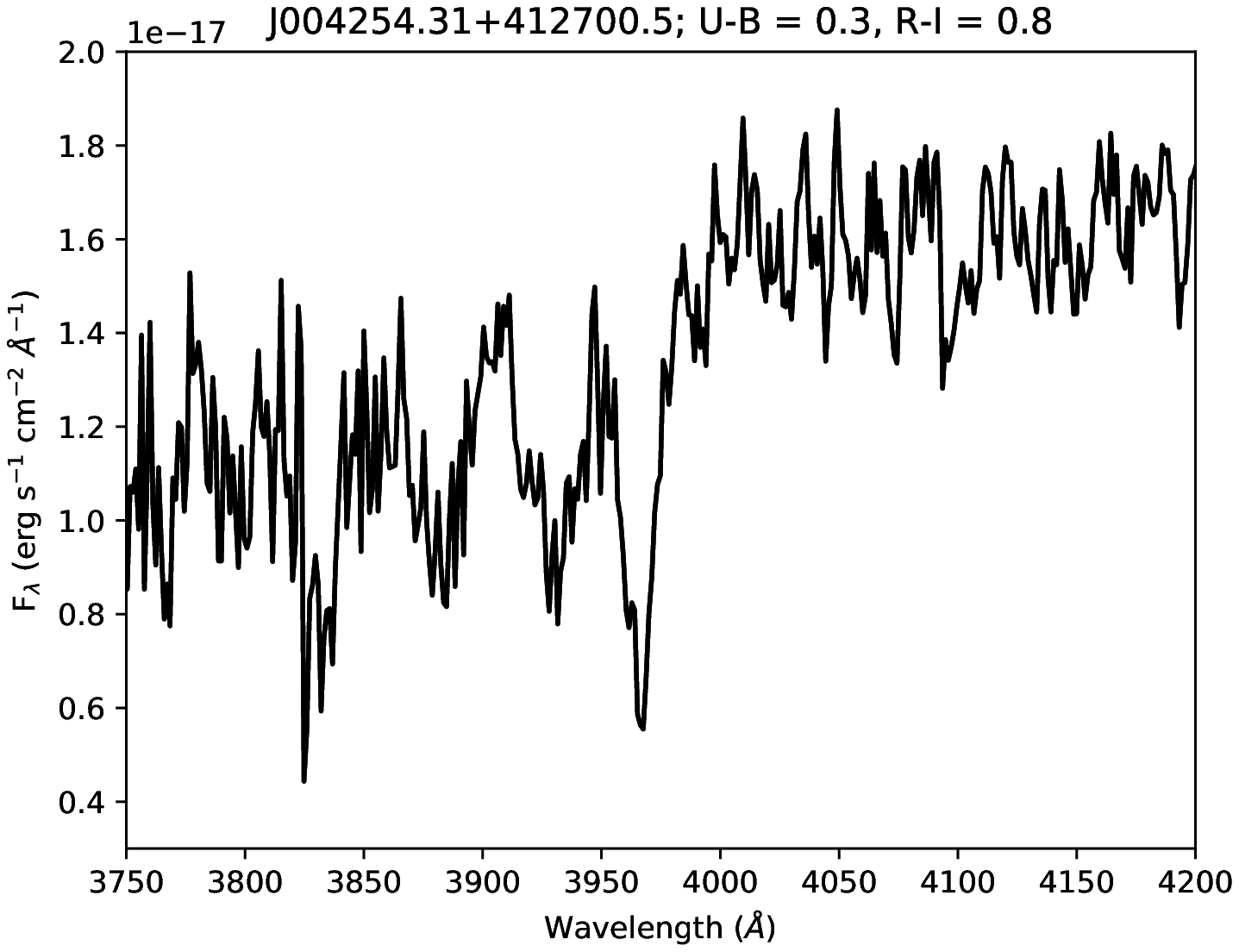}
\plotone{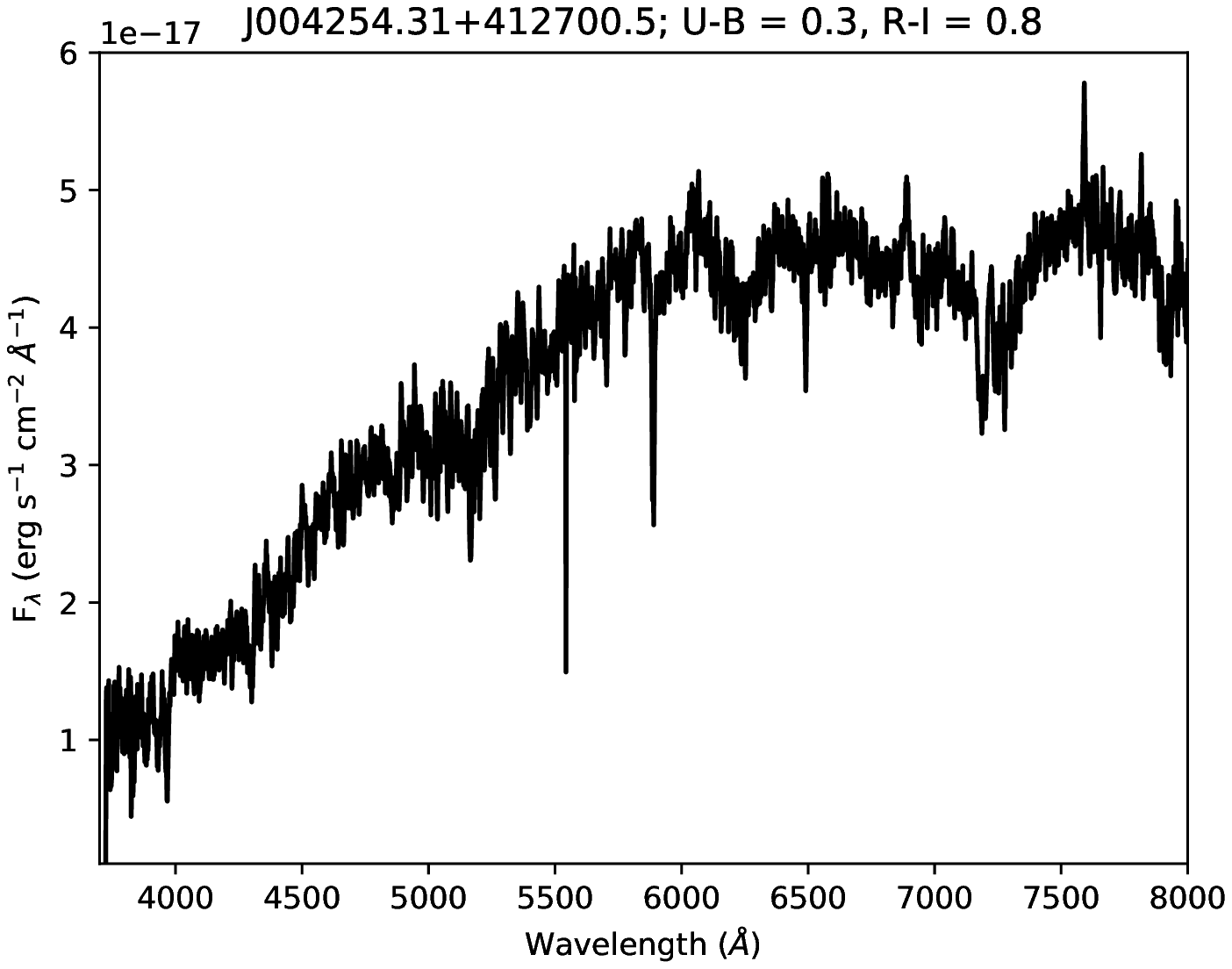}
\caption{\label{fig:specExample} Three representative examples of confirmed RSG+B star binaries. For all three spectra, the entire spectral range (showing both the upper Balmer lines and TiO bands) is shown on the right and an inset of the upper Balmer lines is shown on the left with the Balmer lines marked by red vertical lines. The U-B and R-I values are shown in the title to give a sense of where these stars are located on the color-color plot in Figure~\ref{fig:newDiscovered}. In general, the top star, J003950.51+405307 is located in the upper portion of the triangle. The middle star, J004327.01+412807.7 is located in the bottom left of the triangle and the bottom star, J004254.31+412700.5 is located in the bottom right of the triangle. Notice how the bluer stars have stronger upper Balmer lines while the redder stars have strong TiO bands.}
\end{figure}

Out of the 280 candidates in M31 and M33, 149 were observed. Of these, 63 were confirmed as newly discovered RSG+B star binaries. Their locations in U-B vs.\ R-I color space are plotted in Figure~\ref{fig:newDiscovered} as yellow circles. (It should be noted that because Hectospec is a mult-fiber instrument, flux calibration is inherently difficult because each fiber will have a different transmission curve and hence be dependent upon the accuracy of the flat field correction \citep{Fabricant2008}. Thus, when plotting these stars on a color-color plot we relied on their LGGS colors as opposed to photometry derived from the spectra.) Note that they span the entire range of color space where \citet{NeugentRSG} expected to find RSG binaries. Representative examples of confirmed binary RSGs within the observed color-color space are shown in Figure~\ref{fig:specExample}. 

The coordinates, magnitudes and general classifications of the stars discussed below are listed in Table~\ref{tab:ObsParams}.

\subsection{Contaminants}
While 42\% (63/149) of the observed stars were RSG+B star binary systems, the remaining 58\% were contaminants. Below we detail the types of contaminants including the single RSGs, blue stars, and other interesting stars. Within each section we discuss how to remove such contaminates in future surveys.

\subsubsection{Potentially Single RSGs}
As is shown in Figure~\ref{fig:newDiscovered}, 11\% (17/149) of the observed candidates showed strong TiO bands but no upper Balmer lines. However, a few of them show strong signal in the blue but no upper Balmer lines. This could be due to a few possibilities. The first is that there is a B-type star companion but the SNR simply isn't high enough to detect weak Balmer lines. Another possibility is that we are seeing the effects of dust scattering (similar to what is seen in normal reflection nebula). Many RSGs have extra extinction compared to their OB star neighbors; these stars also show upturns in their UV spectra \citep{Massey05, Levesque05}. As mentioned in \S 1, RSGs have high mass-loss rates. As the mass is expelled, the dust grains absorb the red light while reflecting the blue light \citep{Jura1976}. Thus, some RSGs show what appears to be a blue reflection nebula. This has been seen in such RSGs as VY CMa and $\mu$ Cep \citep{Shenoy2016}. Higher resolution and UV data is needed to either confirm or deny the existence of such a feature.

Given that the majority of these TiO-band spectra show upturns in the blue, we wouldn't exactly call these contaminates. Thus we would rather get additional data to either confirm or deny the presence of a binary companion or reflection nebula as opposed to removing them from our list of candidates.

\subsubsection{Blue Stars}
There were additionally quite a few stars that showed strong Balmer lines but no TiO bands. These stars comprised 28\% (41/149) of the sample. The majority of these stars were B-type stars (28/41 = 68\%) but there were a few A-type stars as well (13/41 = 32\%). Their locations in color-color space are shown in Figure~\ref{fig:newDiscovered} as cyan dots (B stars) and magenta stars (A stars). Note that the majority of them are clustered around the lower-most R-I colors (R-I $\sim$ 0.5) but across a wide range of U-B color-space. Since we are preferentially finding single blue (B and A) stars at low R-I colors, we believe we have adequately sampled that region of color-space for binary RSGs and we would not find any more RSG+B star binaries at lower R-I colors. With the exception of two of the A-type stars, the rest were pushed into our region of interest due to high reddening ($B-V > 0.5$) values as determined from LGGS photometry \citep{LGGS}.

Within a small region of color-color space (at low R-I colors), there are 27 A and B-type stars and only 2 RSG+B star binaries. So, if we are interested in completeness, it is still necessary to sample this region. However, 93\% of this region was contaminated by A and B-type stars. To maximize the number of RSG+B star binaries found from the minimum number of spectroscopic follow-up observations, this region should be sampled last. 

\subsubsection{Other Contaminants}
Although our program was highly successful in identifying RSG with early-type companions, there were invariably some ``losers." Twenty-four spectra fell into this category, an interesting hodgepodge. About half of these are likely foreground stars and are of no further interest here. Two others proved to be RSGs that were located in regions of strong nebulosity. The remainder, however, proved quite interesting in their own right. Twelve objects turn out to be previously unknown background QSOs and galaxies, and will be discussed in a short companion paper (Massey et al., submitted).

Removing this assortment of contaminants is quite difficult since they span the entire range of color-color space. However, some of them could have been removed using the original LGGS \citep{LGGS} images. The galaxies are not quite point sources on the images (although the QSOs are) and thus could have been eliminated by their fuzzy structure. But for the remaining point-sources, there is no easy way of removing them from the list of targets.

\subsubsection{J004032.98+404102.8 - A Symbiotic RSG?}

\begin{figure}
\plotone{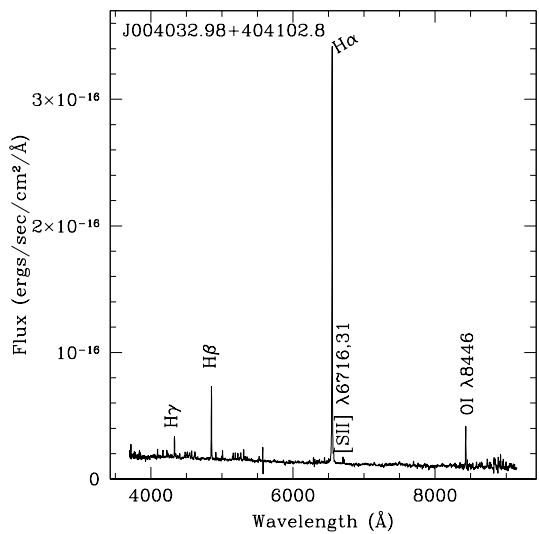}
\caption{\label{fig:weird} The spectrum of J004032.98+404102.8.  The weak, unlabeled lines below 5000\AA\ are predominantly Fe II.}
\end{figure}

One contaminant star, J004032.98+404102.8, stands out in particular. Its spectrum is shown in Figure~\ref{fig:weird}. We find Balmer emission with an extremely strong decrement, along with numerous low-excitation Fe lines, primarily Fe\,{\sc ii}. We were unable to find any forbidden Fe lines, although the nebular lines [S\,{\sc ii}] $\lambda \lambda 6716, 31$ are weakly present, as is the [N\,{\sc ii}] $\lambda 6583$. Very weak [O\,{\sc iii}] $\lambda 5007$ may be present. 

What is the nature of this unusual find?  We are indebted to our colleague Dr.\ Howard Bond for helping us answer this. The object was detected both by the {\it Swift} Ultraviolet / Optical Telescope (UVOT), and by the Wide-field Infrared Survey Explorer (WISE), indicating both hot and cool components. This is a classic description of symbiotic stars providing the first strong evidence that they consisted of a red giant and a white dwarf (see, e.g., \citealt{Boyarchuk}). An inspection of the atlas of symbiotic stars by \citet{SymbioticAtlas} reveals a near identical match between the optical spectrum of V3929 Sgr (their Fig.~75) and that of our M31 star.  

However, there is still a mystery. J004032.98+404102.8 is clearly a member of M31, as its radial velocity (measured via the Balmer lines) is $-543\pm2$ km s$^{-1}$, consistent with that expected given its location in M31's disk, $-531$ km s$^{-1}$ using the assumptions laid out in \citet{MasseySilva}. Thus, the star has a high visual luminosity, $M_V\sim -5.1$. The cool components of symbiotic stars are typically M giants, or at most bright giants (see, e.g., \citealt{Keyes}); these should have absolute visual magnitudes near 0 \citep{AAQ}. A value of $-5$ is more typical of a red {\it super}giant.

Putting together these findings suggest that this may actually be another successful detection of a RSG binary, albeit in an unexpected form. We propose that this object is some sort of high-mass analog of a normal symbiotic star. In a normal symbiotic system, the red giant dumps material onto a white dwarf, and in the process creates an accretion disk. ``D-type" symbiotics usually contain a Mira variable and are surrounded by optically thick dust, obscuring the donor star \citep{Symbiotics}.  Based upon the absolute magnitude, we suspect that this is a RSG+OB companion pair, with the matter transfer creating both the accretion disk and contributing to the obscuration of the donor star. The OB star is exciting some nebula emission, but high obscuration leads to a strong Balmer decrement and the near invisibility of the [O\,{\sc iii}] line. The Fe emission would presumably arise from the accretion disk. It would be worthwhile to monitor this star to see if its spectrum or brightness vary, as one finds with many symbiotics. Although it has been previously identified as a ``miscellaneous" variable (V2285) by \citealt{M31vars}, its amplitude is not much larger than the errors.  

\subsection{Completeness}
The average SNR of our spectra around the upper Balmer lines (3700-4000\AA) was $25\pm9$ per 2\AA\ resolution. \citet{NeugentRSG} found that a SNR of $\sim$100 per 2\AA\ resolution element in the blue was sufficient to detect the upper Balmer lines. Thus, in order to conclusively prove whether the RSG has a B star companion, we needed a much higher SNR than what we obtained. It should also be noted that the spectral resolution of the MMT data was 5\AA\ and thus not adequate for detecting the weakest lined Balmer lines even with a high enough SNR. So, of the 14 RSGs that did not show upper Balmer lines, it cannot be ruled out that they have B star companions with weak upper Balmer lines. Data with a higher SNR and resolution is needed.

We also must consider the likelihood that RSG+B star binaries exist outside of the targeted search area within color-color space. This number is difficult to quantify and many of our confirmed RSG+B star binaries fall on the dividing line as shown in Figure~\ref{fig:newDiscovered}. Considering the increasing contamination of B and A stars near the lower values in R-I, we believe that region of color-color space has been adequately explored. However, we cannot rule out there being more RSG+B star binaries in the redder regions (hypothenuse of the triangle). Sadly, these systems are not rare. They are comprised of RSG + B star binaries where the Balmer lines are quite weak compared to the flux from the RSG. This most commonly occurs with a RSG + BV pair or a binary where the B star has high reddening and thus the Balmer lines have been slightly ``filled in." Given the weakness of the Balmer lines in these spectra, we would need a higher SNR (much closer to $\sim$100, as discussed above) to detect such stars. At this point, the SNR and resolution requirements become almost prohibitive towards large-scale spectroscopic follow up and we may need to devise another method to identify such binaries.

\section{Newly Confirmed Small Magellanic Cloud RSG+B Star Binaries}
In addition to our M31 and M33 observations, we obtained spectra of four RSG+B star binary candidates in the Small Magellanic Cloud (SMC). As was described in \citet{NeugentRSG}, 25 out of 598 previously observed RSGs in the Magellanic Clouds showed evidence of upper Balmer lines (the coordinates and magnitudes of these stars are listed in Table~\ref{tab:ObsParamsMC}). Ten additional RSGs in the sample also showed potential upper Balmer absorption but had SNRs that were too low (around 20 per 2\AA\ resolution) for this to be conclusive. We re-observed four of these low-SNR candidates that were chosen as a representative sample using the 6.5-m Magellan Echellette (MagE) on the Baade Magellan telescope at Las Campanas. We confirmed all four of them as RSG+B star binaries due to the presence of upper Balmer lines. We additionally made a very interesting discovery -- the first known RSG+Be star binary!

\subsection{Observations and Reductions}
All four stars were observed under clear conditions with 0\farcs7 to 1\farcs0 seeing using a $1\arcsec$ slit. Exposure times ranged from 300s (to obtain an unsaturated spectrum of the Be star) to 1800s. MagE provides coverage from 3100\AA\ to 1$\mu$m at a resolving power $R$ of 4100. Thus, we were able to observe the wavelength regions of both the upper Balmer lines and the TiO bands at high resolution. Spectrophotometric standards taken throughout the night allowed for flux calibration before the orders were combined. The data were extracted using a combination of the IRAF echelle package and the `mtools'\footnote{`mtools' is available for download from the Las Campanas Observatory web page.} IRAF routines originally designed by Jack Baldwin for the reduction of MIKE spectra. 

\subsection{The First RSG+Be Star Binary System}
\begin{figure}
\epsscale{0.45}
\plotone{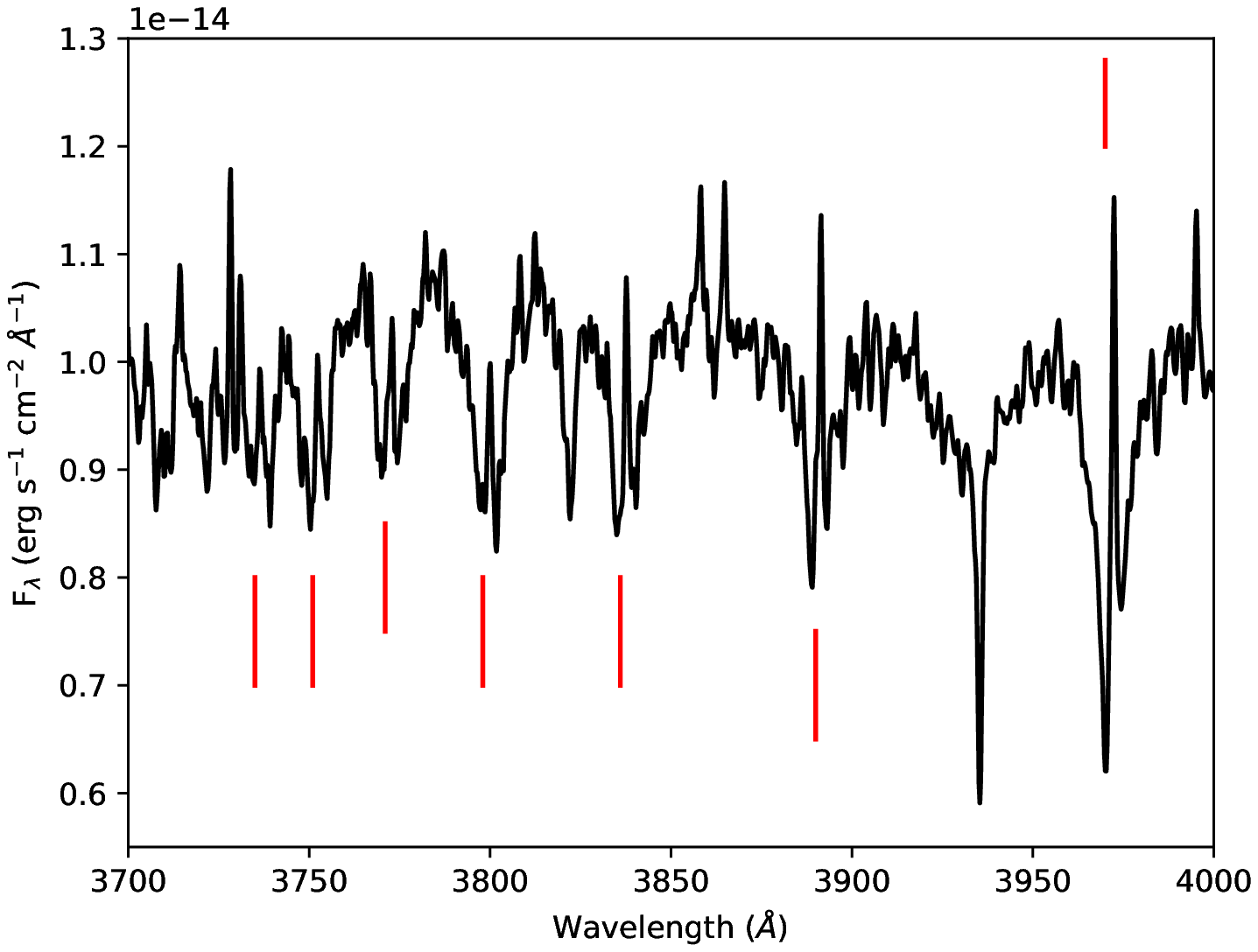}
\plotone{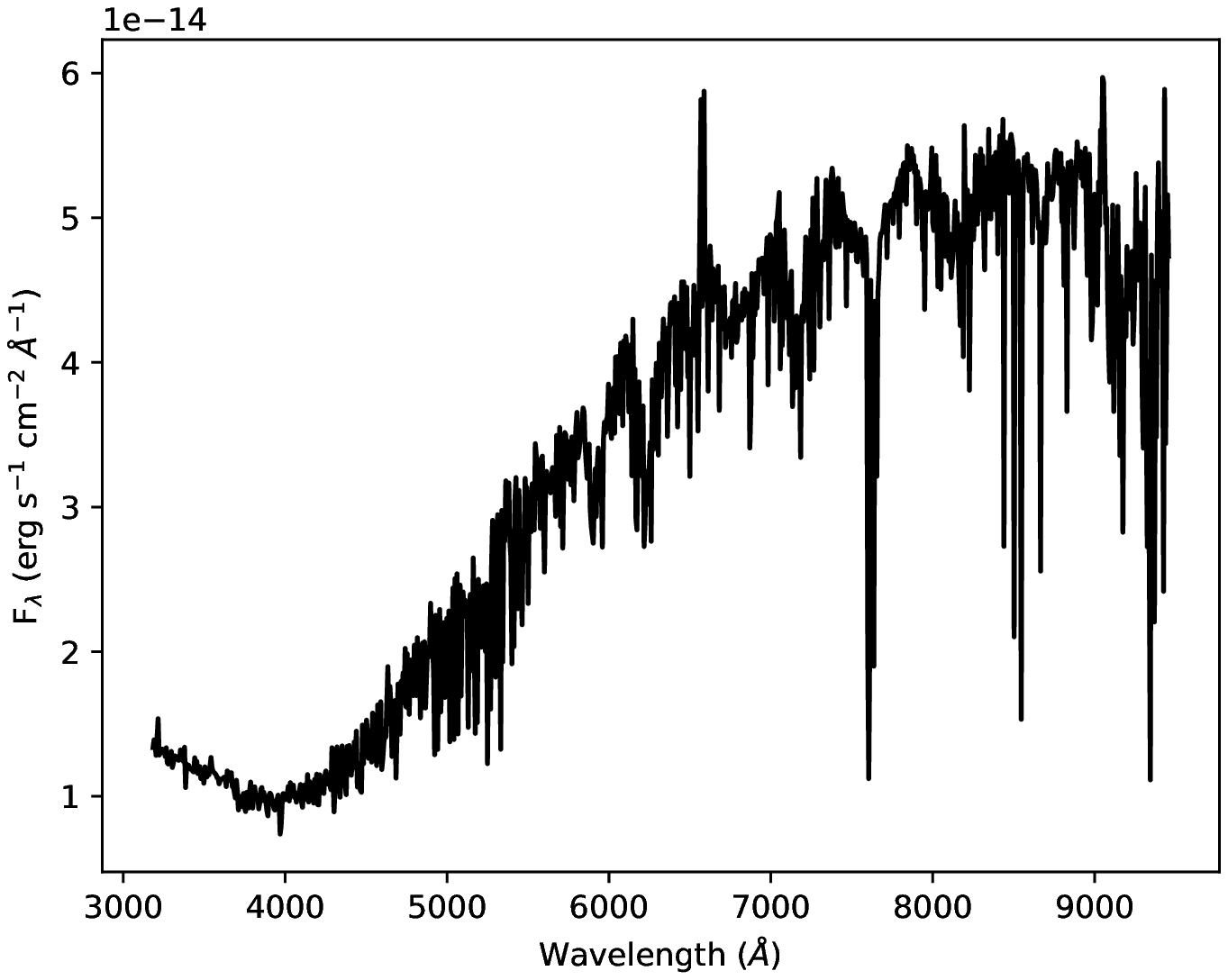}
\caption{\label{fig:BeStar} Spectrum of RSG+Be star binary. The figure on the right shows the entire spectrum including the strong TiO bands coming from the RSG. Also note that H$\alpha$ is in emission near the center of the spectrum. This is indicative of a Be star. The figure on the left shows the emission lines within the cores of the upper Balmer lines. The Balmer lines are marked by red lines.}
\end{figure}

One of the four SMC stars we observed shows strong upper Balmer lines in addition to TiO bands, making it a RSG+B star binary. However, the spectrum also shows emission lines within the core of each upper Balmer line making it different from any of our previously observed binary spectra. Based on these emission lines, we classify the B star component as a Be star. Figure~\ref{fig:BeStar} shows a spectrum of this star including a subset of the upper Balmer lines with the emission and absorption line components. While Be stars themselves are not rare (in the Milky Way, $\sim20$\% of all B stars are thought to be Be stars; see \citealt{Zorec1997}), this is the first known RSG+Be star binary system.

As defined, Be stars are B stars that show emission within their Balmer lines. This emission is thought to form from material being ejected due to a star's rapid rotation. This material then forms an outwardly diffusing gaseous Keplerian disk. Usual rotational velocities of Be star rapid rotators are around 300 km s$^{-1}$. However, based on the widths of the absorption lines of this newly discovered RSG+Be star binary, it has a rotational velocity of around 150 km s$^{-1}$, much less than a normal Be star. One explanation, as suggested by Doug Gies (private communication) is that this is a Be star with a pole-on orientation so that the rotational velocity we're measuring is much smaller than the actual rotational velocity. This has been seen before, but such stars aren't common. One such example is $\chi$ Oph which has a rotational velocity of around 140 km s$^{-1}$ \citep{Tycner2008}. If we take these two systems to be similar, the inclination of the disk to the line of sight is small, less than 18$^\circ$ for the star to not be rotating faster than the breakup velocity.

Another difference from the more traditional Be star is that the emission lines are quite strong and narrow, though this may be due to the small rotational velocity discussed above. Generally they are wider and double-peaked (compare to Figure 1 in \citealt{Chojnowski2018}). However, Be stars are known to be variable on timescales over a few minutes and the emission has even been seen to disappear and reappear over the timescale of decades \citep{Rivinius2013}. Thus, we plan on re-observing this interesting object to see if there are any significant spectral changes over the period of a few years.

Using the strength of the He\,{\sc i} $\lambda$3819 line to the surrounding Balmer lines, we classify this star as an early BV. \citet{Lesh68} expanded the MK spectral type to include subtypes for Be stars. Based on H$\beta$ being in emission, the higher Balmer lines having emission cores, and Fe\,{\sc ii} lines being present but not dominating the spectrum, we classify this Be star as a Be$_3$V. There are other, more recent, classification systems that take into account various shapes of the Balmer lines and polarization, but the \citet{Lesh68} system is still commonly used. In terms of further classifying either the absorption or emission components of the Be star, our spectra does not have a high enough SNR to identify the prominent classification lines such as He\,{\sc i} / Mg\,{\sc ii} $\lambda 4481$. Radial velocity measurements confirm that both the absorption and emission lines have the same radial velocities and thus they are forming in the same region.

\section{Modeling the Spectra}
With the MMT data obtained, we now have 63 confirmed RSG+B star binaries in M31 and M33 as well as 24 confirmed RSG+B star binaries in the SMC and LMC (see \citealt{NeugentRSG} for more information on the Magellanic Cloud binaries) bringing the total number of RSG binaries to 87. While our spectra of these stars are low resolution, they can still be used for basic modeling. Here we describe our efforts to model the RSG component of the binary using the TiO bands in the red and model the B star component of the binary using the upper Balmer lines in the blue.

\subsection{MARCS RSG Modeling}
\begin{figure}
\epsscale{0.45}
\plotone{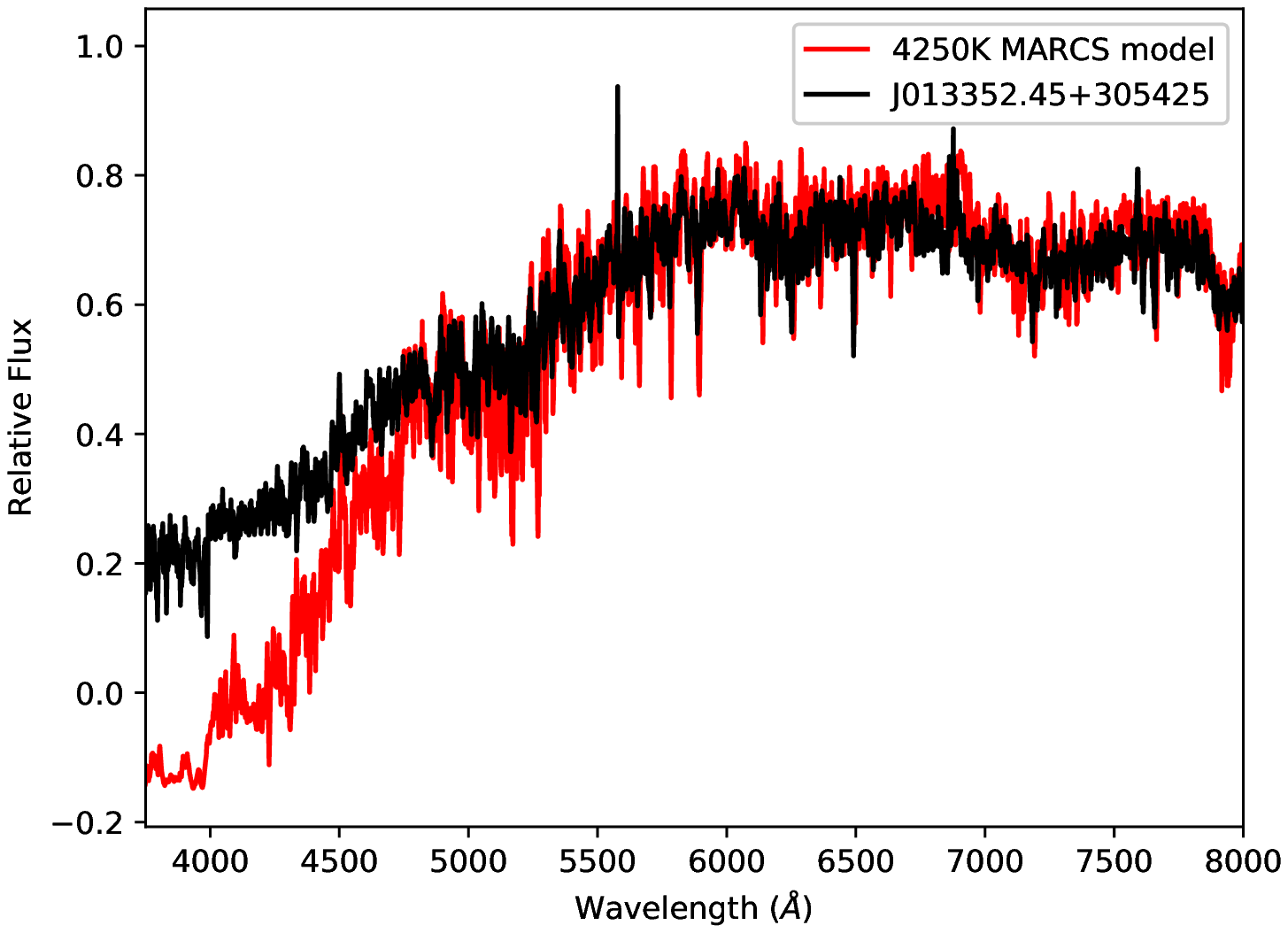}
\plotone{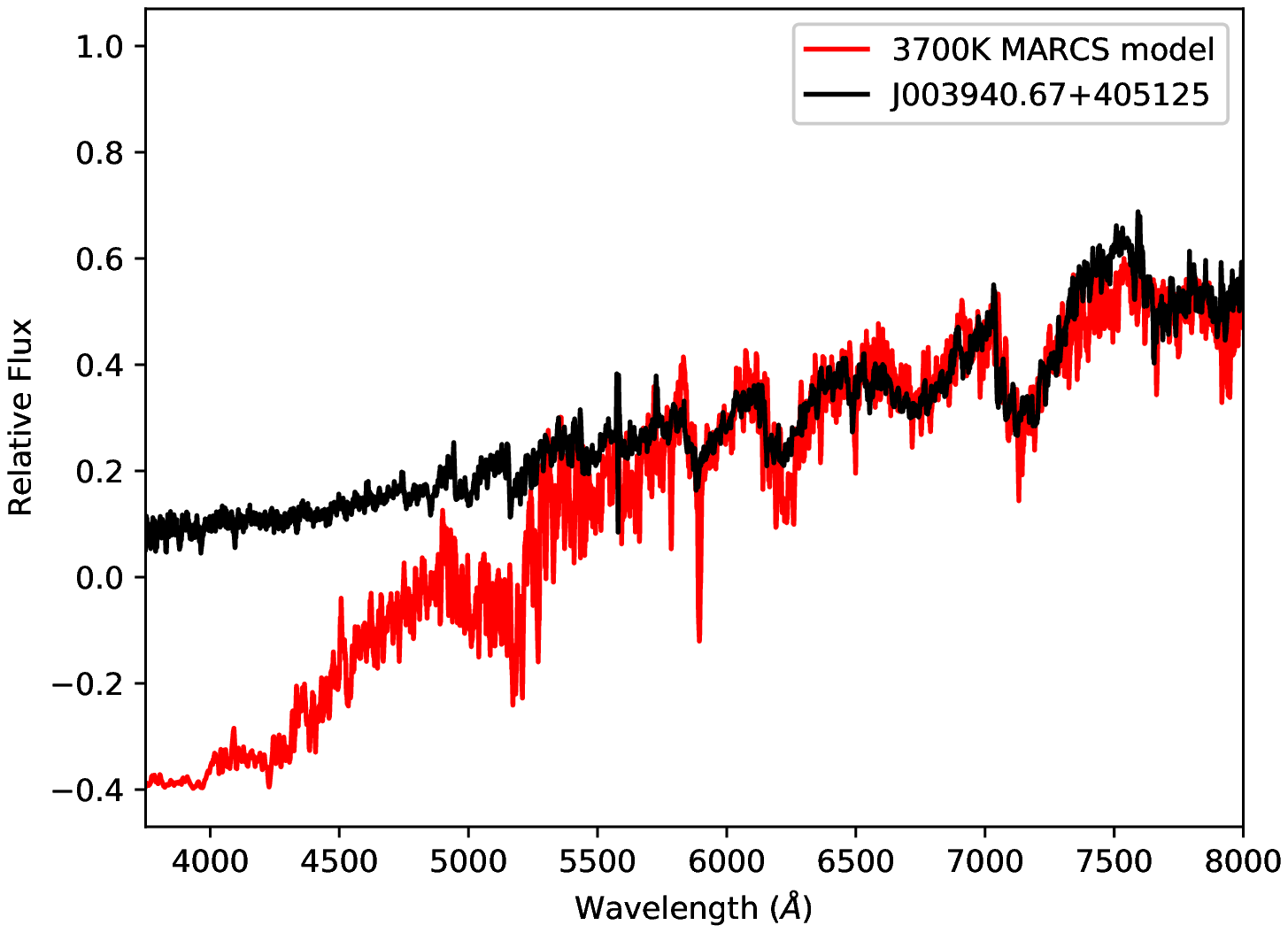}
\caption{\label{fig:MARCSfits} Example fits of MARCS models to observed RSG+B star binary spectra. The figure on the left shows a warm mid-K supergiant spectrum modeled with a 4250\,K MARCS model. Notice the weak TiO bands in contrast to the stronger TiO bands in the figure to the right. The figure on the right shows a cool M supergiant with a 3700\,K MARCS model over plotted. In both cases, the spectra deviate in the blue because of the presence of the B star adding additional flux in the blue.}
\end{figure}

To model the RSG star within the binary system, we used the MARCS models with surface gravities of $\log g = 0.0$, as this is typical of RSGs \citep{Levesque05}. We adopted the solar-metallicity models; the differences in the RSG models between an LMC (0.5 solar) and M31(1.5-2$\times$ solar) are slight; see Figure 5 in \citet{MasseySilva}. We fit the spectral features and continuum to the MARCS models by eye by changing the temperature of models after smoothing the models so they were at a resolution comparable to the observed spectra. As the temperature of the RSG decreases, the strength of the TiO bands increase because it becomes easier for molecules to form at cooler temperatures. Thus, the TiO bands are very prevalent in M stars but the lines can still be used to estimate temperature in mid-Ks, especially the $\lambda 5167$ and $\lambda 6158$ TiO lines. The MARCS models we used to fit the observed spectra ranged from 3000\,K to 4500\,K in increments of 100\,K. Considering the matches we obtain for different temperatures, we estimate our errors are on the order of 50\,K for the M stars and mid-to-late K stars. For the earlier K stars where the TiO bands are the weakest we believe our errors are on the order of 100\,K. This process is very similar to the one used by \citet{Levesque06} when modeling a different set of Magellanic Cloud RSGs. An example best fit model of an early K star and late M star are shown in Figure~\ref{fig:MARCSfits}. Note the increase in the strength of the TiO bands for the later type stars thus allowing for a more precise fit.

\subsection{BSTAR06 B-Star Modeling}
To model the B star component within the binary system we relied on the BSTAR06 models. The strength of the spectral lines depends on the temperature of the star with cooler temperatures producing stronger lines. The width of the spectral lines depends on the luminosity class of the star with dwarfs producing thicker lines than supergiants. Due to the poor resolution of our spectra, the widths of the lines were difficult to accurately measure. It was generally possible to differentiate between a supergiant (very narrow lines) and a dwarf (broad lines) but nearly impossible to distinguish between those two extremes and giants. Thus, degeneracies exist between luminosity classes and temperatures. For example, a star might be adequately fit by both a 19000\,K BV and a 17000\,K BIII. Thus, for this project we eschewed the giant subclass and only considered the two extremes of dwarf vs.\ supergiant. Overall, the temperatures for the giants were 2000\,K cooler than that of the dwarfs and 3000-4000\,K hotter than the supergiants. As is shown in Table~\ref{tab:knownBinaries}, all of the known binaries have BV companions, and thus we believe it is reasonable to preferentially select the BV results over that of a BIII temperature. 

\begin{figure}
\epsscale{1}
\plotone{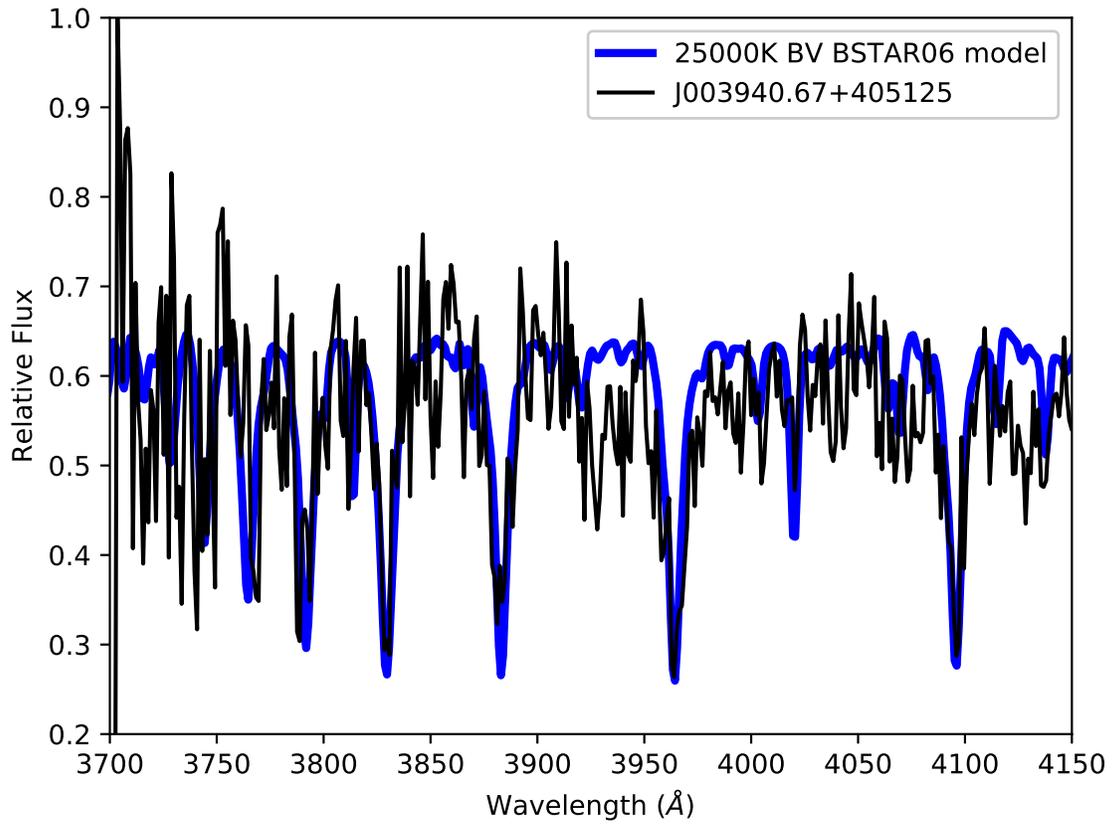}
\caption{\label{fig:BlueFit} Example fit of BSTAR06 model to observed RSG+B star binary spectra. This figure shows a 25000\,K BV model fit to J003940.67+405125. Note the good match between both the strength of the lines (due to the model's temperature) and the width of the lines (due to the model's luminosity class).}
\end{figure}

To fit the Balmer lines of the B stars we first normalized the spectrum using a 5th order cubic spline. We restricted the wavelength range to 3700 - 4200 \AA\ in order to avoid attempting to normalize the strong TiO bands of the RSG. This allowed us to better fit the strengths of the upper Balmer lines of the B star. An example fit is shown in Figure~\ref{fig:BlueFit}. Note that both the strength and width of the lines are well matched by the 25000\,K BV BSTAR06 model.

To double-check our results, we compared the spectral modeling luminosity classes with the absolute magnitude in U, making the assumption that the extinction of these stars was typical of OB stars, and that the flux at U is dominated by the B star component.  We found excellent agreement for the stars we called supergiants, except for the hottest one, J013405.62+304142, whose luminosity is more consistent with a dwarf than a supergiant. For the cooler ``dwarfs," the U-band flux is more consistent with them being giants rather than a dwarfs, but they are not as luminous as supergiants. The warmer dwarfs have U-band luminosities that are consistent with that designation, again indicating that our method worked quite well. There were two notable exceptions, J013254.01+303858 (20,000 K) and J013343.22+303547 (22,000 K), whose U-band luminosities suggest they are supergiants instead, unless the RSG or reflection nebula is adding to the flux.  We have retained the spectroscopic designations but indicated the three discrepant stars with a footnote in Table~\ref{tab:PhysParams}.

\subsection{Spectral Energy Distribution Modeling}
The modeling described above made one key assumption -- that the flux of the binary was unaffected by the B-star component at the TiO bands and was unaffected by the RSG component at the upper Balmer lines. If this isn't the case, it is possible that the B-star flux filled in the TiO bands causing the temperature to be overestimated. Conversely, the flux from the RSG may have filled in the strengths of the upper Balmer lines causing the temperatures to be overestimated as well. The amount the temperature was overestimated will vary from star to star. If the RSG is particularly dusty, the upper Balmer lines will be filled in more than a system with a small amount of dust. The physical separation between the two stars might also alter the overestimate with stars that are closer together being more overestimated than stars that are farther apart. However, there is no easy way to separate the two spectra and thus this is an assumption we've made. 

We originally hoped to model the overall SED of the combined RSG+B star binary. However, after further investigation we determined that the number of free parameters made any resulting fits unreliable. 

The first complication deals with the mass of the individual components. For the RSG, we used a 15$\rm{M_{\odot}}$ supergiant as used in the MARCS models; however, given the 8-25$\rm{M_{\odot}}$ mass range of the RSG population and the effects of the initial mass function this is likely to be an over-estimate for most of the RSGs in our sample. For the B star we used the BSTAR06 models that span a range of masses depending on the type of star (dwarf, giant, supergiant). However, the BSTAR06 models do not take into account that a giant might be 10$\rm{M_{\odot}}$ or even 60$\rm{M_{\odot}}$ but still have the same 20,000\,K temperature. Changing the mass of the B star will drastically change the shape of the SED. While it is possible to estimate the temperature and spectral type of the B star using the line depths and widths (as discussed above), it is not possible to use this method to estimate the mass of the star. Thus, we are unable to accurately describe the SED without knowing the mass of the B star. While further observations (both spectroscopic and photometric) would hypothetically allow us to determine the masses of these stars, this is well outside the scope of this current project.

The other complication is the reddening of the system. From the observed spectra alone we know very little about the dust surrounding the system. We can make the basic assumption that the reddening around the B star is either equal to or less than the reddening around the RSG however it is not possible to determine the ratio. This value will additionally change based on the distance between the two stars (a B star in a close binary system should have a higher reddening than a B star in a wide binary system). While we would ideally like to place constraints on the $A_V$s of the systems, the poor flux calibration stemming from multi-fiber spectra does not allow for any estimates. To learn more about the reddening of both components, and thus the dust production of the RSG, we would need additional data in both the UV (for the B star) and the NIR (for the RSG).

\section{Discussion}
Table~\ref{tab:PhysParams} contains an overview of observational and modeling results for each system including the adopted temperatures and luminosity classes for the B-type stars and temperatures for the RSGs.

\subsection{Overall Trends}
All of the previously known (Galactic) RSG binaries have B dwarfs as their companions (Table I), but here we have discovered RSG with B supergiant companions as well.   We expect B supergiant companions to be rarer given their shorter lifetimes, and our sample of newly discovered RSG+B pairs is a factor of $\sim$9 times larger than all of the previously known RSG binaries. We find that 66\% of our newly identified RSG sample contain B dwarfs, and 34\% contain a B supergiant. Note that, as discussed above, we were not able to distinguish between a dwarf vs.\ giant or giant vs.\ supergiant so we've simplified the classifications into just dwarfs and supergiants. There does not seem to be a correlation between temperature or subtype for either the B stars or the RSGs. The average temperature for the B star (for both dwarf and supergiant) is 21000\,K with a standard deviation of 4000\,K. The average temperature for the RSG (both when paired with a dwarf or supergiant) is 3900\,K with a standard deviation of 150\,K. These values are reasonable for both RSGs and B-type stars. We would expect to see some change in RSG and B star temperatures from galaxy to galaxy as the metallicity changes but small number statistics prevent us from seeing this trend.
	
\subsection{Variability}
RSGs are known to be variable on the timescales of days all the way up to years \citep{Szczygiel10}. Almost all are variable on the order of 1 magnitude or higher while some pulsate between 80 to 3500 days \citep{GuoLi}. Convective cells and hot spots transverse their surface adding additional photometric variations to any observed light curve \citep{Chiavassa11, Baron14, Stothers10}. Additionally, episodic mass loss events cause non-periodic outbursts such as what occurred with V838 Mon in 2002 \citep{Tylenda11}. These variations all add to the complexity of searching for binary companions using light curve variations as a starting point.

Despite these difficulties, we investigated the variability of LMC and SMC RSG+B star binaries using the All-Sky Automated Survey for Supernovae (ASAS-SN) \citep{Shappee2014, ASAS-SN}. Out of the 24 confirmed binaries, five of them were previously identified as variable by ASAS-SN but none of them showed periodic variability. Three additional stars were variable but had not been catalogued as variable by ASAS-SN and the remaining 15 did not appear to vary significantly (less than 0.05 mag) over a period of 1000 days. We were a little surprised by this outcome. The known RSG binaries have periods on the order of years and thus with a baseline of 1000 days we expected to see some indication of variation due to binarity. However, this was not the case. In the future we plan on investigating this issue further by searching for small variations within the periods that might indicate binarity and extending this search to the M31 and M33 binaries using iPTF data but that is outside the scope of this current paper. 

\section{Conclusions and Future Work}
As part of this work we've now observationally shown that it is possible to select RSG+B star binary candidates using photometry alone. Using such photometry we have confirmed 87 new RSG+B star binaries in the Local Group galaxies of M31, M33, the LMC and the SMC. We additionally discovered the first known RSG+Be binary system. Using the BSTAR06 and MARCS model atmosphere codes we estimated temperatures and luminosity classes for the B stars as well as temperatures for the RSGs and found that there is no correlation between luminosity class and temperature for either RSGs or B stars.  

Our next step is to understand the overall binary fraction of RSGs. To do this we need to obtain a statistically significant sample of both single and binary RSGs within a galaxy. Determining this within our own Galaxy is difficult given reddening and our location within the disk, but the galaxies of the Local Group provide excellent test beds for such research. As part of the ongoing work with collaborators, we are in the process of cataloging RSGs within M31, M33, the LMC, SMC and several of the smaller Local Group galaxies. By applying the photometric selection criteria discussed in this paper and obtaining more spectroscopic follow-up, we can detect a statistically significant sample of binary RSGs within each of these Local Group galaxies. This will allow us to put a number on the binary fraction of RSGs and compare it to the binary fraction of other massive stars. We can then compare our results with evolutionary models allowing for further refinements and advances in massive star research.

\acknowledgements
We first thank Doug Gies for his help deciphering the RSG+Be star spectrum. We additionally thank the support staff at both the MMT and Las Campanas for their assistance. We appreciated the anonymous referee's comments which improved this paper. This paper uses data products produced by the OIR Telescope Data Center, supported by the Smithsonian Astrophysical Observatory as well as data gathered with the 6.5 m Magellan telescopes located at Las Campanas Observatory, Chile. This work was supported in part by NSF IGERT grant DGE-1258485 as well as by a fellowship from the Alfred P.\ Sloan Foundation. P.M.'s work was supported by the National Science Foundation under grant AST-1612874.

	
\bibliographystyle{apj}
\bibliography{masterbib}

\begin{deluxetable}{l l l}
\tablecaption{\label{tab:knownBinaries} Previously Known RSG+B Star Binaries}
\tablewidth{0pt}
\tablehead{
\colhead{Name}
&\colhead{Type}
&\colhead{Reference}
}
\startdata
22 Vul & G4I + B9V & \cite{Ahmad1985}\\
31 Cyg & K4Iab + B4IV-V & \cite{Hansen1944, Stencel1984}\\
32 Cyg & K5Iab + B7V & \cite{McLaughlin1950, Chun1992} \\
AL Vel & RSG + BV & \cite{Kilkenny1995}\\
$\beta$ Per & K0IV + B8V & \cite{Sarna1993}\\
BD +59$^\circ$224 & K4.5Ib + B3V & \cite{Gray2004}\\
XX Per & M4Ib + B7V & \cite{Stothers1971}\\
V766 Cen & RSG + BV & \cite{Wittkowski2017}\\
V838 Mon & MI + BV & \cite{Munari2007, Tylenda2009}\\
VV Cep & M2Iab + B0-2V & \cite{Cowley1969, Pollmann2017}\\
$\zeta$ Aur & K5II + B7V & \cite{Harper2016, Wright1970} \\
\enddata
\end{deluxetable}

\clearpage
\begin{deluxetable}{l l l l l r l l}
\tabletypesize{\scriptsize}
\tablecaption{\label{tab:ObsParams} Coordinates and Magnitudes of Classified M31 and M33 Stars\tablenotemark{*}}
\tablewidth{0pt}
\tablehead{
\colhead{LGGS ID}
& \colhead{$\alpha_{\rm 2000}$} 
& \colhead{$\delta_{\rm 2000}$} 
& \colhead{$V$}
& \colhead{$B-V$}
& \colhead{$U-B$}
& \colhead{$R-I$}
& \colhead{Classification}
}
\startdata
J003940.67+405125.4 & 00 39 40.66 & +40 51 25.3 & 19.089 & 1.849 & -0.122 & 1.161 & RSG+B binary\\
J003942.08+405221.4 & 00 39 42.07 & +40 52 21.3 & 18.880 & 0.536 & -0.576 & 0.547 & RSG+B binary\\
J003950.51+405307.4 & 00 39 50.50 & +40 53 07.3 & 18.985 & 0.485 & -0.803 & 1.252 & RSG+B binary\\
J003955.78+410426.7 & 00 39 55.77 & +41 04 26.6 & 19.847 & 0.602 &  0.257 & 0.672 & B-type star\\
J004021.42+404035.7 & 00 40 21.41 & +40 40 35.6 & 19.685 & 0.812 & -0.396 & 0.622 & RSG+B binary\\
J004024.49+405418.0 & 00 40 24.48 & +40 54 17.9 & 19.978 & 0.526 & -0.539 & 0.468 & A-type star\\
J004024.87+404900.9 & 00 40 24.86 & +40 49 00.8 & 19.999 & 1.250 & -0.241 & 0.575 & A-type star\\
J004030.12+404502.3 & 00 40 30.11 & +40 45 02.2 & 19.916 & 0.423 & -0.744 & 0.821 & RSG+B binary\\
J004030.95+403923.5 & 00 40 30.94 & +40 39 23.4 & 18.587 & 0.871 & -0.016 & 0.630 & B-type star\\
J004032.98+404102.8 & 00 40 32.97 & +40 41 02.7 & 19.702 & 0.238 & -0.914 & 0.409 & high-mass symbiotic\\
J004033.98+404429.9 & 00 40 33.97 & +40 44 29.8 & 18.963 & 0.998 &  0.002 & 0.738 & RSG+B binary\\
J004034.80+404328.6 & 00 40 34.79 & +40 43 28.5 & 19.644 & 0.701 & -0.554 & 0.632 & B-type star\\
J004036.31+405058.0 & 00 40 36.30 & +40 50 57.9 & 19.922 & 0.996 &  0.013 & 0.567 & B-type star\\
J004043.78+404335.3 & 00 40 43.77 & +40 43 35.2 & 19.830 & 0.186 & -0.866 & 0.442 & B-type star\\
J004053.01+404356.3 & 00 40 53.00 & +40 43 56.2 & 19.738 & 0.778 & -0.165 & 0.531 & B-type star\\
J004059.48+410229.4 & 00 40 59.47 & +41 02 29.3 & 19.234 & 0.827 & -0.499 & 1.105 & RSG+B binary\\
J004100.50+410432.5 & 00 41 00.49 & +41 04 32.4 & 19.917 & 0.726 & -0.279 & 0.516 & B-type star\\
J004101.95+410911.4 & 00 41 01.94 & +41 09 11.3 & 19.878 & 1.456 &  0.520 & 0.692 & RSG\\
J004106.51+404848.8 & 00 41 06.50 & +40 48 48.7 & 19.950 & 1.131 & -0.418 & 1.003 & RSG+B binary\\
J004115.88+404011.4 & 00 41 15.87 & +40 40 11.3 & 19.680 & 1.407 & -0.221 & 0.910 & B-type star\\
J004117.43+410837.5 & 00 41 17.42 & +41 08 37.4 & 19.346 & 1.215 & -0.466 & 1.131 & RSG+B binary\\
J004134.79+411418.1 & 00 41 34.78 & +41 14 18.0 & 19.907 & 1.076 & -0.037 & 0.666 & RSG+B binary\\
J004136.62+410018.3 & 00 41 36.61 & +41 00 18.2 & 19.831 & 1.454 &  0.519 & 0.601 & RSG\\
J004139.49+404033.4 & 00 41 39.48 & +40 40 33.3 & 19.482 & 0.335 & -0.379 & 0.501 & A-type star\\
J004143.13+411734.9 & 00 41 43.12 & +41 17 34.8 & 19.907 & 0.701 & -0.207 & 0.563 & B-type star\\
J004145.25+411229.4 & 00 41 45.24 & +41 12 29.3 & 19.446 & 0.288 & -0.074 & 0.532 & RSG+B binary\\
J004149.73+405300.1 & 00 41 49.72 & +40 53 00.0 & 19.537 & 0.509 & -0.492 & 0.892 & RSG+B binary\\
J004151.42+411804.7 & 00 41 51.41 & +41 18 04.6 & 19.992 & 0.776 & -0.755 & 0.919 & RSG+B binary\\
J004153.05+405235.7 & 00 41 53.04 & +40 52 35.6 & 19.714 & 0.495 &  0.244 & 0.649 & A-type star\\
J004158.62+405338.6 & 00 41 58.61 & +40 53 38.5 & 19.440 & 0.713 & -0.767 & 1.153 & RSG+B binary\\
J004158.87+405316.7 & 00 41 58.86 & +40 53 16.6 & 18.416 & 0.743 & -0.184 & 0.532 & B-type star\\
J004204.77+411340.8 & 00 42 04.76 & +41 13 40.7 & 19.799 & 1.158 &  0.518 & 0.720 & RSG\\
J004205.41+412826.0 & 00 42 05.40 & +41 28 25.9 & 18.700 & 0.137 & -0.482 & 0.484 & B-type star\\
J004205.57+413345.8 & 00 42 05.56 & +41 33 45.7 & 19.633 & 0.880 & -0.363 & 0.694 & RSG+B binary\\
J004209.69+411340.7 & 00 42 09.68 & +41 13 40.6 & 19.605 & 0.517 &  0.208 & 0.693 & RSG+B binary\\
J004211.64+412341.2 & 00 42 11.63 & +41 23 41.1 & 19.944 & 0.809 &  0.098 & 0.807 & A-type star\\
J004217.21+411743.2 & 00 42 17.20 & +41 17 43.1 & 18.421 & 0.533 & -0.152 & 0.501 & A-type star\\
J004229.89+412034.6 & 00 42 29.88 & +41 20 34.5 & 19.913 & 0.218 & -0.790 & 0.509 & B-type star\\
J004238.73+405349.9 & 00 42 38.72 & +40 53 49.8 & 19.692 & 1.234 & -0.143 & 1.073 & RSG+B binary\\
J004239.82+412054.7 & 00 42 39.81 & +41 20 54.6 & 19.730 & 1.246 &  0.714 & 0.613 & RSG\\
J004241.27+413156.8 & 00 42 41.26 & +41 31 56.7 & 19.397 & 0.883 &  0.055 & 0.601 & A-type star\\
J004243.97+411344.9 & 00 42 43.96 & +41 13 44.8 & 19.583 & 0.511 &  0.218 & 0.773 & RSG+HII\\
J004246.20+411232.8 & 00 42 46.19 & +41 12 32.7 & 19.692 & 0.385 & -0.625 & 0.709 & RSG\\
J004246.22+411309.1 & 00 42 46.21 & +41 13 09.0 & 19.908 & 0.642 &  0.389 & 0.800 & RSG\\
J004247.66+405810.6 & 00 42 47.65 & +40 58 10.5 & 19.262 & 0.576 &  0.354 & 0.624 & A-type star\\
J004253.28+411246.4 & 00 42 53.27 & +41 12 46.3 & 19.673 & 0.336 &  0.118 & 0.587 & RSG+B binary\\
J004253.29+411010.8 & 00 42 53.28 & +41 10 10.7 & 19.898 & 0.960 & -0.173 & 0.768 & RSG+B binary\\
J004254.31+412700.5 & 00 42 54.30 & +41 27 00.4 & 19.268 & 1.439 &  0.337 & 0.788 & RSG+B binary\\
J004256.79+410230.5 & 00 42 56.78 & +41 02 30.4 & 19.878 & 0.755 & -0.452 & 0.512 & B-type star\\
J004259.35+410211.0 & 00 42 59.34 & +41 02 10.9 & 19.619 & 0.899 & -0.115 & 0.573 & B-type star\\
J004313.97+411119.7 & 00 43 13.96 & +41 11 19.6 & 19.167 & 0.688 & -0.441 & 0.938 & A-type star\\
J004314.06+410844.9 & 00 43 14.05 & +41 08 44.8 & 19.485 & 0.523 &  0.260 & 0.739 & A-type star\\
J004327.01+412808.7 & 00 43 27.00 & +41 28 08.6 & 17.668 & 0.224 & -0.599 & 0.665 & RSG+B binary\\
J004328.81+411854.9 & 00 43 28.80 & +41 18 54.8 & 19.377 & 0.807 & -0.713 & 0.811 & RSG+B binary\\
J004332.50+412205.7 & 00 43 32.49 & +41 22 05.6 & 19.708 & 0.361 &  0.136 & 0.572 & B-type star\\
J004338.56+412511.6 & 00 43 38.55 & +41 25 11.5 & 19.936 & 1.107 &  0.355 & 0.610 & RSG\\
J004339.23+412443.5 & 00 43 39.22 & +41 24 43.4 & 17.809 & 1.201 &  0.502 & 0.605 & B-type star\\
J004344.53+411104.6 & 00 43 44.52 & +41 11 04.5 & 19.753 & 0.996 &  0.042 & 0.934 & RSG+B binary\\
J004345.11+410946.7 & 00 43 45.10 & +41 09 46.6 & 19.952 & 0.542 &  0.283 & 0.571 & A-type star\\
J004345.75+411451.7 & 00 43 45.74 & +41 14 51.6 & 19.829 & 1.086 &  0.202 & 0.663 & B-type star\\
J004348.91+411629.2 & 00 43 48.90 & +41 16 29.1 & 19.794 & 0.793 & -0.354 & 0.573 & RSG+B binary\\
J004351.83+411431.7 & 00 43 51.82 & +41 14 31.6 & 19.337 & 0.759 & -0.326 & 0.566 & RSG+B binary\\
J004352.80+410957.9 & 00 43 52.79 & +41 09 57.8 & 18.098 & 1.219 & -0.532 & 0.915 & B-type star\\
J004359.63+411853.8 & 00 43 59.62 & +41 18 53.7 & 19.804 & 0.798 & -0.249 & 0.802 & RSG\\
J004405.80+411937.4 & 00 44 05.79 & +41 19 37.3 & 19.726 & 0.691 & -0.181 & 0.821 & RSG+B binary\\
J004410.28+411757.6 & 00 44 10.27 & +41 17 57.5 & 19.570 & 1.460 &  0.309 & 0.847 & RSG+B binary\\
J004428.15+411002.0 & 00 44 28.14 & +41 10 01.9 & 19.731 & 0.428 &  0.185 & 0.676 & RSG+B binary\\
J013209.23+302614.8 & 01 32 09.20 & +30 26 14.7 & 19.873 & 1.760 &  0.401 & 0.703 & RSG\\
J013224.33+303155.5 & 01 32 24.30 & +30 31 55.4 & 19.547 & 1.808 & -0.437 & 0.865 & RSG\\
J013232.81+302855.7 & 01 32 32.78 & +30 28 55.6 & 19.794 & 0.705 & -0.523 & 0.631 & B-type star\\
J013250.80+303507.6 & 01 32 50.77 & +30 35 07.5 & 19.940 & 0.468 & -1.047 & 0.792 & RSG+B binary\\
J013253.33+303943.7 & 01 32 53.30 & +30 39 43.6 & 19.889 & 1.643 &  0.467 & 0.797 & RSG\\
J013254.01+303858.0 & 01 32 53.98 & +30 38 57.9 & 18.649 & 0.448 & -0.534 & 0.656 & RSG+B binary\\
J013254.95+302411.1 & 01 32 54.92 & +30 24 11.0 & 19.150 & 0.589 & -0.719 & 0.879 & RSG+B binary\\
J013256.23+302752.1 & 01 32 56.20 & +30 27 52.0 & 19.293 & 1.012 & -0.521 & 0.847 & RSG+B binary\\
J013256.84+302347.9 & 01 32 56.81 & +30 23 47.8 & 19.327 & 0.712 & -0.338 & 0.658 & RSG+B binary\\
J013259.82+303231.5 & 01 32 59.79 & +30 32 31.4 & 19.799 & 1.306 & -0.528 & 0.829 & RSG+B binary\\
J013301.02+303500.7 & 01 33 00.99 & +30 35 00.6 & 19.108 & 1.238 & -0.321 & 1.066 & RSG+B binary\\
J013303.54+303201.2 & 01 33 03.51 & +30 32 01.1 & 18.878 & 2.177 & -0.528 & 1.405 & RSG\\
J013306.83+303039.5 & 01 33 06.80 & +30 30 39.4 & 19.600 & 1.357 & -0.179 & 1.061 & RSG+B binary\\
J013309.52+303410.7 & 01 33 09.49 & +30 34 10.6 & 19.963 & 0.206 &  0.041 & 0.560 & A-type star\\
J013312.20+304849.4 & 01 33 12.17 & +30 48 49.3 & 18.871 & 2.097 & -0.625 & 1.054 & RSG\\
J013314.45+302615.2 & 01 33 14.42 & +30 26 15.1 & 19.832 & 1.232 & -0.262 & 0.891 & RSG+B binary\\
J013315.30+302248.3 & 01 33 15.27 & +30 22 48.2 & 19.514 & 1.032 & -0.451 & 0.796 & RSG+B binary\\
J013320.17+304655.0 & 01 33 20.14 & +30 46 54.9 & 19.827 & 1.302 &  0.026 & 0.764 & RSG+B binary\\
J013323.19+304016.2 & 01 33 23.16 & +30 40 16.1 & 19.937 & 0.711 & -0.236 & 0.749 & RSG+B binary\\
J013323.64+302836.8 & 01 33 23.61 & +30 28 36.7 & 19.905 & 0.742 & -0.590 & 0.767 & RSG+B binary\\
J013324.56+304908.3 & 01 33 24.53 & +30 49 08.2 & 19.923 & 1.022 & -0.063 & 0.999 & RSG+B binary\\
J013328.78+303002.5 & 01 33 28.75 & +30 30 02.4 & 19.889 & 0.290 & -0.442 & 0.527 & B-type star\\
J013339.51+304316.6 & 01 33 39.48 & +30 43 16.5 & 18.875 & 0.382 & -0.878 & 1.200 & RSG+B binary\\
J013342.45+304237.2 & 01 33 42.42 & +30 42 37.1 & 19.890 & 0.952 & -0.136 & 0.892 & RSG+B binary\\
J013342.55+305309.6 & 01 33 42.52 & +30 53 09.5 & 18.985 & 0.750 & -0.880 & 1.117 & RSG\\
J013343.22+303547.9 & 01 33 43.19 & +30 35 47.8 & 18.750 & 0.411 & -0.945 & 0.936 & RSG+B binary\\
J013345.91+303915.6 & 01 33 45.88 & +30 39 15.5 & 19.357 & 0.113 & -0.757 & 0.680 & B-type star\\
J013347.51+304154.9 & 01 33 47.48 & +30 41 54.8 & 19.972 & 1.769 & -0.708 & 1.418 & RSG+B binary\\
J013347.82+304324.9 & 01 33 47.79 & +30 43 24.8 & 18.436 & 0.237 & -0.421 & 0.564 & RSG+B binary\\
J013349.94+302928.8 & 01 33 49.91 & +30 29 28.7 & 18.102 & 0.983 & -0.842 & 1.098 & RSG+B binary\\
J013351.13+303922.0 & 01 33 51.10 & +30 39 21.9 & 19.897 & 0.242 & -0.797 & 0.481 & B-type star\\
J013352.45+305425.6 & 01 33 52.42 & +30 54 25.5 & 18.830 & 1.466 &  0.280 & 0.723 & RSG+B binary\\
J013355.62+304116.6 & 01 33 55.59 & +30 41 16.5 & 19.916 & 1.217 & -0.236 & 1.147 & RSG+B binary\\
J013356.09+303834.5 & 01 33 56.06 & +30 38 34.4 & 19.376 & 0.686 & -0.613 & 1.257 & RSG+B binary\\
J013358.33+302216.5 & 01 33 58.30 & +30 22 16.4 & 19.698 & 1.138 & -0.173 & 0.744 & RSG+B binary\\
J013358.53+304754.7 & 01 33 58.50 & +30 47 54.6 & 19.971 & 0.923 & -0.506 & 0.921 & RSG+B binary\\
J013359.54+303200.4 & 01 33 59.51 & +30 32 00.3 & 19.733 & 0.575 & -0.480 & 1.083 & RSG+B binary\\
J013359.67+304940.2 & 01 33 59.64 & +30 49 40.1 & 19.869 & 0.109 & -0.526 & 0.486 & B-type star\\
J013401.67+304050.7 & 01 34 01.64 & +30 40 50.6 & 19.935 & 1.687 & -0.301 & 1.045 & RSG+B binary\\
J013402.58+304310.1 & 01 34 02.55 & +30 43 10.0 & 19.265 & 0.613 & -0.733 & 1.002 & RSG+B binary\\
J013403.54+304201.9 & 01 34 03.51 & +30 42 01.8 & 19.414 & 0.883 & -0.194 & 0.512 & B-type star\\
J013403.85+303911.0 & 01 34 03.82 & +30 39 10.9 & 19.965 & 1.477 &  0.367 & 0.708 & B-type star\\
J013404.04+304804.7 & 01 34 04.01 & +30 48 04.6 & 17.817 & 1.246 &  0.355 & 0.639 & B-type star\\
J013404.56+302459.9 & 01 34 04.53 & +30 24 59.8 & 19.820 & 0.780 & -0.022 & 0.748 & RSG+B binary\\
J013405.62+304142.8 & 01 34 05.59 & +30 41 42.7 & 19.994 & 0.855 & -0.275 & 0.809 & RSG+B binary\\
J013408.10+304616.3 & 01 34 08.07 & +30 46 16.2 & 19.971 & 0.875 &  0.020 & 0.650 & RSG+B binary\\
J013408.17+304813.6 & 01 34 08.14 & +30 48 13.5 & 19.719 & 0.308 &  0.047 & 0.647 & A-type star\\
J013409.66+304721.9 & 01 34 09.63 & +30 47 21.8 & 19.578 & 0.569 &  0.266 & 0.596 & B-type star\\
J013413.98+302759.0 & 01 34 13.95 & +30 27 58.9 & 18.592 & 1.034 & -0.188 & 1.134 & RSG+B binary\\
J013413.99+305035.5 & 01 34 13.96 & +30 50 35.4 & 19.888 & 0.386 &  0.081 & 0.530 & B-type star\\
J013414.05+310450.4 & 01 34 14.02 & +31 04 50.3 & 19.856 & 1.294 &  0.336 & 0.819 & RSG\\
J013415.32+303804.3 & 01 34 15.29 & +30 38 04.2 & 18.916 & 0.756 & -0.776 & 1.238 & RSG+B binary\\
J013424.81+310456.4 & 01 34 24.78 & +31 04 56.3 & 19.045 & 1.867 &  0.155 & 0.879 & RSG\\
J013431.25+304538.8 & 01 34 31.22 & +30 45 38.7 & 19.023 & 0.216 & -1.022 & 0.980 & RSG+B binary\\
J013431.85+305400.8 & 01 34 31.82 & +30 54 00.7 & 19.525 & 0.498 & -0.645 & 1.038 & RSG+B binary\\
J013436.97+304514.9 & 01 34 36.94 & +30 45 14.8 & 19.786 & 1.273 & -0.415 & 1.284 & RSG+B binary\\
J013439.43+304551.4 & 01 34 39.40 & +30 45 51.3 & 19.193 & 1.596 & -0.520 & 1.391 & RSG+B binary\\
J013448.91+303140.4 & 01 34 48.88 & +30 31 40.3 & 19.961 & 0.990 & -0.277 & 0.967 & RSG+B binary\\
\enddata
\tablenotetext{*}{All photometry from \citealt{LGGS}.}
\end{deluxetable}

\clearpage
\begin{deluxetable}{l l l l r r}
\tabletypesize{\scriptsize}
\tablecaption{\label{tab:ObsParamsMC} Coordinates and Magnitudes of Confirmed SMC and LMC RSG+B Star Binaries\tablenotemark{*}}
\tablewidth{0pt}
\tablehead{
\colhead{ID}
& \colhead{$\alpha_{\rm 2000}$} 
& \colhead{$\delta_{\rm 2000}$} 
& \colhead{$V$}
& \colhead{$U-B$}
& \colhead{$V-B$}
}
\startdata  
J00464984-7313525 & 00 46 49.84 & -73 13 52.5 & 13.599 & 0.043 & -1.313\\
J00513280-7205493 & 00 51 32.80 & -72 05 49.3 & 13.096 & -0.061 & -1.238\\
J00522647-7245159 & 00 52 26.47 & -72 45 15.9 & 12.496 & -0.341 & -1.348\\
J00531861-7242074 & 00 53 18.61 & -72 42 07.4 & 13.486 & 0.549 & -1.518\\
J00532528-7215376 & 00 53 25.28 & -72 15 37.6 & 13.296 & 0.549 & -1.448\\
J00532655-7212033 & 00 53 26.55 & -72 12 03.3 & 13.700 & 0.847 & -1.355\\
J00534451-7233192 & 00 53 44.51 & -72 33 19.2 & 13.226 & 0.099 & -1.238\\
J00535739-7313335 & 00 53 57.39 & -73 13 33.5 & 13.879 & 0.193 & -1.432\\
J00543483-7229512 & 00 54 34.83 & -72 29 51.2 & 13.732 & 1.549 & -1.754\\
J01014357-7238252 & 01 01 43.57 & -72 38 25.2 & 13.076 & -0.041 & -1.288\\
J01023794-7235547 & 01 02 37.94 & -72 35 54.7 & 13.376 & 0.109 & -1.198\\
J01024076-7217173 & 01 02 40.76 & -72 17 17.3 & 13.516 & -0.071 & -0.438\\
J01033730-7158448 & 01 03 37.30 & -71 58 44.8 & 13.096 & 0.529 & -1.388\\
J04595731-6748133 & 04 59 57.31 & -67 48 13.3 & 13.176 & -0.211 & -0.888\\
J05030232-6847203 & 05 03 02.32 & -68 47 20.3 & 13.612 & -0.133 & -1.097\\
J05043378-6806235 & 05 04 33.78 & -68 06 23.5 & 13.536 & -0.71 & -0.754\\
J05065284-6841123 & 05 06 52.84 & -68 41 12.3 & 14.469 & -0.659 & -0.707\\
J05083180-6853372 & 05 08 31.80 & -68 53 37.2 & 14.253 & 0.152 & -1.594\\
J05263040-6948033 & 05 26 30.40 & -69 48 03.3 & 14.683 & 1.9 & -1.402\\
J05274747-6913205 & 05 27 47.47 & -69 13 20.5 & 12.736 & 0.229 & -1.528\\
J05275113-6910460 & 05 27 51.13 & -69 10 46.0 & 12.596 & 0.319 & -1.398\\
J05292143-6900202 & 05 29 21.43 & -69 00 20.2 & 12.016 & -0.711 & -0.548\\
J05300119-6956382 & 05 30 01.19 & -69 56 38.2 & 13.406 & 0.655 & -1.222\\
J05353280-6904191 & 05 35 32.80 & -69 04 19.1 & 13.126 & 0.299 & -1.508\\
\enddata
\tablenotetext{*}{SMC photometry from \citealt{ZaritskySMC} and LMC photometry from \citealt{ZaritskyLMC}.}
\end{deluxetable}

\clearpage
\begin{deluxetable}{l l l l}
\tabletypesize{\scriptsize}
\tablecaption{\label{tab:PhysParams} Modeled Physical Parameters of Confirmed RSG+B Star Binaries}
\tablewidth{0pt}
\tablehead{
\colhead{ID}
& \colhead{B-star $T_{\rm eff}$ (K)}
& \colhead{B-star luminosity class}
& \colhead{RSG $T_{\rm eff}$ (K)}
}
\startdata
J003940.67+405125.4 & 19000 & V & 3700 \\
J003942.08+405221.4 & 17000 & V & 4000 \\
J003950.51+405307.4 & 17000 & I & 3900 \\
J004021.42+404035.7 & 16000 & V & 4000 \\
J004030.12+404502.3 & 20000 & V & 3900 \\
J004033.98+404429.9 & 17000 & V & 4000 \\
J004059.48+410229.4 & 18000 & I & 3900 \\
J004106.51+404848.8 & 26000 & V & 4000 \\
J004117.43+410837.5 & 21000 & V & 3800 \\
J004134.79+411418.1 & 16000 & I & 4250 \\
J004145.25+411229.4 & 16000 & V & 4000 \\
J004149.73+405300.1 & 20000 & V & 4000 \\
J004151.42+411804.7 & 16000 & I & 3900 \\
J004158.62+405338.6 & 22000 & V & 3800 \\
J004205.57+413345.8 & 21000 & V & 4000 \\
J004209.69+411340.7 & 18000 & I & 4250 \\
J004238.73+405349.9 & 25000 & V & 4000 \\
J004253.28+411246.4 & 21000 & V & 4000 \\
J004253.29+411010.8 & 26000 & V & 4000 \\
J004254.31+412700.5 & 30000 & I & 4000 \\
J004328.81+411854.9 & 30000 & I & 3900 \\
J004344.53+411104.6 & 21000 & I & 4250 \\
J004348.91+411629.2 & 30000 & I & 4250 \\
J004351.83+411431.7 & 24000 & I & 4250 \\
J004405.80+411937.4 & 26000 & V & 4100 \\
J004410.28+411757.9 & 30000 & I\tablenotemark{a} & 4000 \\
J004428.15+411002.0 & 27000 & V & 4000 \\
J004649.84-731352. & 17000 & V & 4250 \\
J005132.80-720549 & 25000 & V & 3800 \\
J005226.47-724515 & 23000 & I & 3800 \\
J005318.61-724207 & 17000 & V & 4000 \\
J005325.28-721537 & 19000 & V & 3800 \\
J005326.55-721203 & 22000 & V & 3800 \\
J005344.51-723319 & 17000 & I & 3900 \\
J005357.39-731333 & 17000 & V & 3900 \\
J005434.83-722951 & 20000 & V & 3800 \\
J010143.57-723825 & 17000 & I & 4250 \\
J010237.94-723554 & 20000 & V & 4250 \\
J010240.76-721717 & 19000 & V & 3900 \\
J010337.30-715844 & 19000 & V & 3900 \\
J013250.80+303507.6 & 30000 & I & 4250 \\
J013254.01+303858.0 & 20000 & V\tablenotemark{b} & 3900 \\
J013254.95+302411.1 & 18000 & I & 3800 \\
J013256.23+302752.1 & 24000 & I & 3900 \\
J013256.84+302347.9 & 22000 & V & 4000 \\
J013259.82+303231.5 & 30000 & I & 4250 \\
J013301.02+303500.7 & 30000 & I & 3900 \\
J013306.83+303039.5 & 26000 & I & 3800 \\
J013314.45+302615.2 & 20000 & V & 3700 \\
J013315.30+302248.3 & 28000 & V & 4000 \\
J013320.17+304655.0 & 28000 & I & 4100 \\
J013323.19+304016.2 & 24000 & V & 4000 \\
J013323.64+302836.8 & 21000 & V & 4000 \\
J013324.56+304908.3 & 26000 & V & 3700 \\
J013339.51+304316.6 & 24000 & I & 3700 \\
J013342.45+304237.2 & 23000 & V & 3900 \\
J013343.22+303547.9 & 22000 & V\tablenotemark{b} & 3900 \\
J013347.51+304154.9 & 18000 & I & 3800 \\
J013347.82+304324.9 & 20000 & V & 4000 \\
J013349.94+302928.8 & 27000 & I & 3800 \\
J013352.45+305425.6 & 27000 & V & 4250 \\
J013355.62+304116.6 & 29000 & V & 3800 \\
J013356.09+303834.5 & 20000 & V & 3800 \\
J013358.33+302216.5 & 15000 & V & 4000 \\
J013358.53+304754.7 & 17000 & I & 3700 \\
J013359.54+303200.4 & 30000 & V & 3800 \\
J013401.67+304050.7 & 21000 & I & 3800 \\
J013402.58+304310.1 & 25000 & V & 3800 \\
J013405.62+304142.8 & 27000 & I & 3900 \\
J013408.10+304616.3 & 21000 & I & 4000 \\
J013413.98+302759.0 & 19000 & V & 3700 \\
J013415.32+303804.3 & 30000 & V & 3700 \\
J013431.25+304538.8 & 17000 & I & 3800 \\
J013431.85+305400.8 & 20000 & I & 3700 \\
J013436.97+304514.9 & 21000 & I & 3700 \\
J013439.43+304551.4 & 18000 & I & 3700 \\
J013448.91+303140.4 & 25000 & I & 3800 \\
J045957.31-674813 & 16000 & V & 3700 \\
J050302.32-684720 & 16000 & V & 4000 \\
J050433.78-680623 & 18000 & V & 3900 \\
J050831.80-685337 & 17000 & V & 3700 \\
J052630.40-694803 & 15000 & V & 3700 \\
J052747.47-691320 & 17000 & I & 3600 \\
J052751.13-691046 & 23000 & V & 3700 \\
J052921.43-690020 & 20000 & V & 3800 \\
J053001.19-695638 & 17000 & V & 3800 \\
J053532.80-690419 & 20000 & V & 3400 \\
\enddata
\tablenotetext{a}{U-band luminosity suggests V.}
\tablenotetext{b}{U-band luminosity suggests I.}
\end{deluxetable}

\end{document}